\documentclass[%
 reprint,
superscriptaddress,
 amsmath,amssymb,
 aps,
prb,
]{revtex4-1}

\usepackage{graphicx}
\usepackage{epstopdf}
\usepackage{dcolumn}
\usepackage{bm}
\usepackage{gensymb}
\usepackage{upgreek}
\usepackage{hyperref}

\begin{document}

\preprint{APS/123-QED}

\title{Pressure-Driven Valence Increase and Metallization in Kondo Insulator Ce$_3$Bi$_4$Pt$_3$}

\author{Daniel J. Campbell}
\affiliation{Center for Nanophysics and Advanced Materials, Department of Physics, University of Maryland, College Park, Maryland 20742, USA}
\author{Zachary E. Brubaker}
\affiliation{Lawrence Livermore National Laboratory, Livermore, California 94550, USA}
\affiliation{Department of Physics, University of California, Davis, California 95616, USA}
\author{Connor Roncaioli}
\affiliation{Center for Nanophysics and Advanced Materials, Department of Physics, University of Maryland, College Park, Maryland 20742, USA}
\author{Prathum Saraf}
\affiliation{Center for Nanophysics and Advanced Materials, Department of Physics, University of Maryland, College Park, Maryland 20742, USA}
\author{Yuming Xiao}
\affiliation{HPCAT, X-ray Science Division, Argonne National Laboratory, Argonne, Illinois 60439, USA}
\author{Paul Chow}
\affiliation{HPCAT, X-ray Science Division, Argonne National Laboratory, Argonne, Illinois 60439, USA}
\author{Curtis Kenney-Benson}
\affiliation{HPCAT, X-ray Science Division, Argonne National Laboratory, Argonne, Illinois 60439, USA}
\author{Dmitry Popov}
\affiliation{HPCAT, X-ray Science Division, Argonne National Laboratory, Argonne, Illinois 60439, USA}
\author{Rena J. Zieve}
\affiliation{Department of Physics, University of California, Davis, California 95616, USA}
\author{Jason R. Jeffries}
\affiliation{Lawrence Livermore National Laboratory, Livermore, California 94550, USA}
\author{Johnpierre Paglione}
\affiliation{Center for Nanophysics and Advanced Materials, Department of Physics, University of Maryland, College Park, Maryland 20742, USA}

\date{\today}

\begin{abstract}

We report the results of high pressure x-ray diffraction, x-ray absorption, and electrical transport measurements of Kondo insulator Ce$_3$Bi$_4$Pt$_3$ up to 42~GPa, the highest pressure reached in the study of any Ce-based KI. We observe a smooth decrease in volume and movement toward intermediate Ce valence with pressure, both of which point to increased electron correlations. Despite this, temperature-dependent resistance data show the suppression of the interaction-driven ambient pressure insulating ground state. We also discuss potential ramifications of these results for the predicted topological KI state.

\end{abstract}

\maketitle

\section{\label{sec:Intro}Introduction}

Kondo insulators (KIs) are materials for which an energy gap at the Fermi level arises not simply from basic charge transfer considerations, but instead due to more complex hybridization between conduction electrons and outer valence shell magnetic electrons\cite{PietrusLaDoping}. Ce$_3$Bi$_4$Pt$_3$, with a Kondo gap of about 10~meV, was one of the first KIs to be identified\cite{SeveringInelasticNeutron, PietrusLaDoping}. While preliminary work has suggested a trivial nature\cite{WakehamTKIs,CaoTheory}, it has been predicted, along with fellow KI SmB$_6$, to harbor topological surface states\cite{DzeroTKIs}.

Because the properties of KIs are determined in large part by electron-electron interactions, they can be especially sensitive to changes in local environment. Previous studies have shown that it is possible to close the Kondo gap in Ce$_3$Bi$_4$Pt$_3$ with high magnetic fields or elemental substitution, leading to a correlated metallic ground state\cite{HundleyMR, JaimePulsedField, PietrusLaDoping, BoebingerMRPulsedField}. Pressures below 10~GPa close the hybridization gap in other KIs, such as SmB$_6$\cite{BeilleSmB6Gap}, CeNiSn\cite{KurisuCeNiSn}, and CeRhSb\cite{UwatokoCeRhSb}. That is in spite of the generic expectation that unit cell compression should increase that hybridization between 4\textit{f} and conduction electrons that is responsible for the Kondo gap. A previous transport study of Ce$_3$Bi$_4$Pt$_3$ claimed that by 15~GPa the gap increased to several times its ambient pressure value\cite{CooleyCe343Pressure}. However, work with other KIs has shown that pressure-induced changes to atomic valence, the Fermi level, or the \textit{k}-space configuration of the gap can override this and lead to metallic behavior.

To further explore the evolution of Ce$_3$Bi$_4$Pt$_3$ with high pressure, we have carried out x-ray diffraction (XRD), x-ray absorption near-edge structure (XANES), and electrical resistance measurements using diamond anvil cells (DACs). A maximum pressure of 42~GPa was reached in the x-ray studies, which to our knowledge is the highest pressure attained in the study of any Ce-based KI. Transport measurements show a Kondo-related feature that increases in temperature with pressure, while at the same time the resistance increase with cooling becomes smaller and changes form. In contrast to the previous study, we identify this as a signal of metallization of the material, rather than evidence for a robust insulator. The unit cell compresses in a manner well-described by a standard equation of state, and the Ce valence increases from its ambient pressure value of 3.09 to about 3.3. All together, the three experiments point to a smooth crossover to a more metallic ground state at high pressure.

\section{\label{sec:Methods}Experimental Details}

Single crystal samples of Ce$_3$Bi$_4$Pt$_3$ were grown with the typical Bi flux method\cite{HundleyCe343}. Ce, Pt, and Bi were combined in a 1:1:20 ratio in an alumina crucible, which was then sealed in a quartz tube filled with partial Ar atmosphere, heated to 1150~$\degree{}$C at a rate of 50~$\degree{}$C/hour, and held at that temperature for five hours. The growth was then cooled at 2~$\degree{}$C/h to 520~$\degree{}$C, where it was quickly removed from the furnace and spun in a centrifuge to separate crystals from excess molten Bi. Ambient pressure powder XRD outside of a DAC was done with a Rigaku MiniFlex~600. XRD and XANES measurements were performed at room temperature at beamlines 16-IDB and 16-IDD, respectively, of the Advanced Photon Source (APS) at Argonne National Laboratory, in coordination with the High Pressure Collaborative Access Team (HPCAT). XRD data were converted to one dimensional raw data using the DIOPTAS software\cite{PrescherDioptas} and refined using GSAS-II\cite{TobyGSAS-II}. Electrical transport measurements were made in 9~T and 16~T Quantum Design Physical Properties Measurement Systems down to a base temperature of 2~K.

Diamond anvil cells were used to generate pressure and prepared slightly differently for each experiment. The XRD DAC had a rhenium gasket and used neon as a pressure medium. Re is a hard metal, which makes it easier to reach high pressures, and Ne, like the other noble gases, is very hydrostatic\cite{KlotzPressureMedia}. XANES measurements require x-rays emitted by the sample to travel through the gasket, thus it is necessary to choose a material with a low atomic number and lower absorption. For this reason we used beryllium, though it is much softer in comparison to Re and carries a higher risk of toxicity. Mineral oil was used as a pressure medium, because while it is less hydrostatic it simplifies loading the sample into the potentially hazardous Be gasket. Resistance measurements were done in a ``designer DAC'' with tungsten contacts embedded in the diamond\cite{PattersonDesignerDAC,WeirDesignerDAC}. The gasket was made of the nonmagnetic cobalt-nickel alloy MP35N, and insulating steatite was the pressure medium. This material, a solid, is much more liable to pressure inhomogeneities than the other two media used in this study. However, transport measurements in the designer DAC require the sample to be touching the W pads when pressure is first applied, and so a liquid or gas pressure medium would not work.

Copper powder was placed in the XRD cell and refined jointly with Ce$_3$Bi$_4$Pt$_3$, and the well-established pressure dependence of the Cu lattice parameter was used to calculate pressure\cite{DewaeleCuEOS}. For XANES and transport measurements, pressure was calibrated using the known pressure dependence of the fluorescence lines of ruby spheres placed in the cells\cite{PiermariniRubyCalibration}. The wavelength of the fluorescence was noted before and after each measurement, and we present the average value of the two with error bars representing the maximum and minimum observed pressure. In the case of resistance measurements, two ruby spheres were placed in different parts of the gasket hole, as the solid pressure medium is likely to result in greater pressure gradients. Values for both were averaged as measured at room temperature before and after temperature cycling, since differing thermal contraction of different components of the DAC can lead to pressure changes. Up to 20~GPa, the pressure reading of the two rubies did not differ by more than 2~GPa, though disagreement was larger at pressures beyond that value. For transport and XANES, samples were cleaved from larger single crystals, while XRD was done with a ground powder of single crystals.

\section{\label{sec:XRD}X-ray Diffraction}

\begin{figure}
    \centering
    \includegraphics[width=0.38\textwidth]{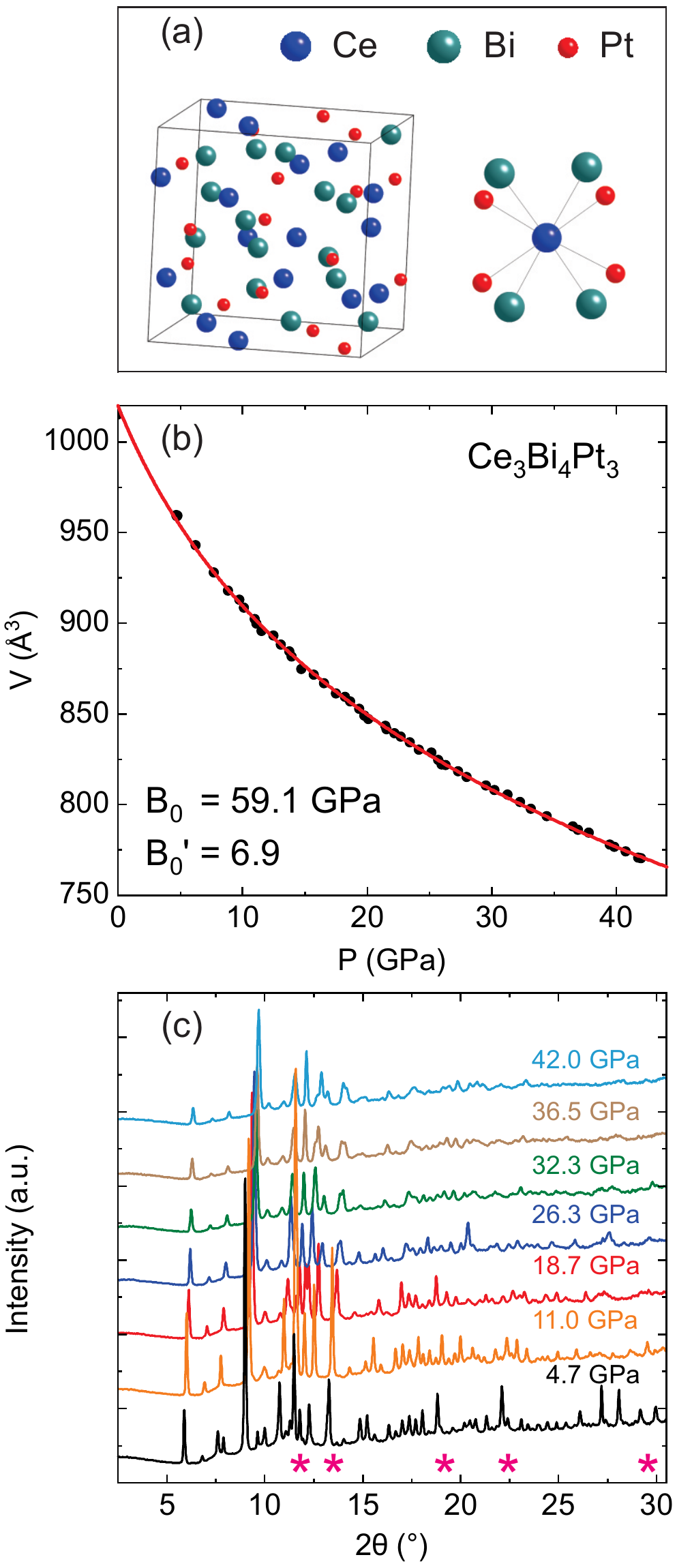}
    \caption{(a) The Ce$_3$Bi$_4$Pt$_3$ unit cell, along with the fundamental Ce units (tetrahedrally coordinated to both Bi and Pt) that compose the larger structure. (b) Volume change with pressure of Ce$_3$Bi$_4$Pt$_3$. The red line is a fit to the third-order Birch-Murnaghan equation of state [Eq.~1]. The 0~GPa volume was determined outside of a DAC. (c) Diffraction patterns (offset) for various pressures taken at room temperature. Cu peaks are marked with pink asterisks.}
    \label{fig:Figure1}
\end{figure}

Ce$_3$Bi$_4$Pt$_3$ forms in the cubic \textit{I$\overline{\textit{4}}$3d} space group (no.~220) at ambient pressure [Fig.~1(a)]. Our samples were found to have an ambient pressure lattice parameter \textit{a}~=~10.05~\AA{}, close to the reference value\cite{HundleyCe343}. It has been noted before that the unit cell is smaller than expected based on interpolation between the trivalent La and Pr equivalents, a sign of ambient pressure intermediate valence\cite{FiskKondoInsulators}. XRD measurements were made at room temperature between 4.7 and 42.0~GPa at the APS with 30~keV (0.4133~\AA) radiation. Ce$_3$Bi$_4$Pt$_3$ and Cu were refined simultaneously in the pattern. The third-order Birch-Murnaghan equation of state is\cite{BirchEOS}

\small
\begin{equation} 
 P(V) = \frac{3B_0}{2}\big[(\frac{V_0}{V})^\frac{7}{3} - (\frac{V_0}{V})^\frac{5}{3}\big]\Big\{1 + \frac{3}{4}(B_0'-4)\big[(\frac{V_0}{V})^\frac{2}{3} - 1\big]\Big\}
\end{equation} 
\normalsize

where $B_0$ is the bulk modulus and $B_0'$ its derivative. Pressure was determined by applying this equation to the refined Cu lattice parameter using reference values (133~GPa and 5.01, respectively)\cite{DewaeleCuEOS}. The Ce$_3$Bi$_4$Pt$_3$ lattice parameter also shows a smooth decrease with pressure and the volume change fits well to the same equation of state form [Fig.~1(b)], yielding $B_0 = 59.1$~GPa and $B_0' = 6.9$. $B_0$ is lower than the value of about 95~GPa (at 300~K) previously determined from thermal expansion measurements\cite{KweiThermalExpansion} at ambient pressure and 1.77~GPa. The data never differ from the fit by more than 0.07\%; however, the refinements in the 10-20~GPa region had a slightly larger error, attributable primarily to distorted peak shapes. This may indicate the threshold to a more noticeable change in interaction strength and Ce valence, a notion we will expand on in the next section. As evident in Fig.~1(c), there is no qualitative change to the diffraction pattern that would indicate a structural transition, except for the typical loss of intensity with rising pressure. 

Of interest when considering Kondo interactions is the Ce-Ce nearest neighbor distance. The unit cell can be viewed as being made up of Ce atoms separately tetrahedrally coordinated with Bi and Pt [Fig.~1(a)]. Each Ce atom has eight Ce nearest neighbors with which it shares both a Bi and Pt atom. The Ce-Ce spacing decreases from 4.70~\AA{} at ambient pressure to 4.288~\AA{} at 42.0~GPa, an 8.8\% change. The high pressure value is still much larger than in bulk Ce, for which room temperature distances are 3.649~\AA{} and 3.429~\AA{} in the trivalent $\gamma$ and collapsed, higher valence $\alpha$ phases, respectively\cite{GschneidnerREMetals}. But the $\gamma$-$\alpha$ transition constitutes only about a 6.0\% decrease in Ce-Ce distance. CeRu$_4$Sb$_{12}$ transitions from metallic to semiconducting by 10~GPa with a Ce-Ce distance, initially nearly twice as large as in Ce$_3$Bi$_4$Pt$_3$, that changes by only about 3\%.\cite{KuritaCeRu4Sb12} Therefore the pressure range of our study should be well within the range of influencing physical properties.

\section{\label{sec:XAS}Cerium Valence Measurements}

A cerium atom can easily lose its 5\textit{d} and two 6\textit{s} electrons. It has only a single 4\textit{f} electron, which is not as tightly bound as in the rare earths with more complete \textit{f} shells. As a result, Ce can easily be found in a 3+ (4$f^1$) or 4+ (4$f^0$) valence configuration. Interaction with conduction electrons can delocalize the remaining electron in space and favor the 4+ state. Thus there is an inherent link between Ce valence above three and electron correlations. In CeRhSb and CeNiSn, signatures of intermediate-valent behavior disappear upon replacement of Ce by La, Zr, or Ti, which all have empty \textit{f} shells\cite{SlebarskiCeZrTiNiSn, SlebarskiCeImpurities}. The same occurs in Ce$_3$Bi$_4$Pt$_3$ doped with trivalent La, which suppresses the resistance increase\cite{HundleyCe343}. Pressure can increase valence, not only in the case of Ce but other rare earths such as Sm and Yb\cite{YamaokaCeTIn5Valence, JosephCePXAS, RueffCeCu2Si2Pressure, ButchSmB6Valence, FuseYbValence, BrubakerYb114}.

The valence change of Ce in Ce$_3$Bi$_4$Pt$_3$ with pressure was quantified through x-ray absorption measurements at the L$_3$  edge (denoting the 2$p_{3/2} \rightarrow{}$5\textit{d} transition), done in partial fluorescence yield mode. XANES is a two step process: incoming radiation promotes a core level electron into an unoccupied state in the conduction band. Another core electron then drops to a lower energy level to fill the newly created hole, emitting a photon in the process. The energy required to excite the initial electron depends on the screening of others surrounding it. In the case of Ce, this is affected by the configuration of the 4\textit{f} electron. By sending in energy-tuned radiation in the region around the Ce absorption ``edges'' and tracking the number of emitted photons it is possible to infer the average 4\textit{f} occupation\cite{BrubakerCeRhIn5Valence, ButchSmB6Valence}. This is done by weighing relative peak heights at different input energies corresponding to different valence states\cite{ButchSmB6Valence}.

\begin{figure*}
    \centering
    \includegraphics[width=1\textwidth]{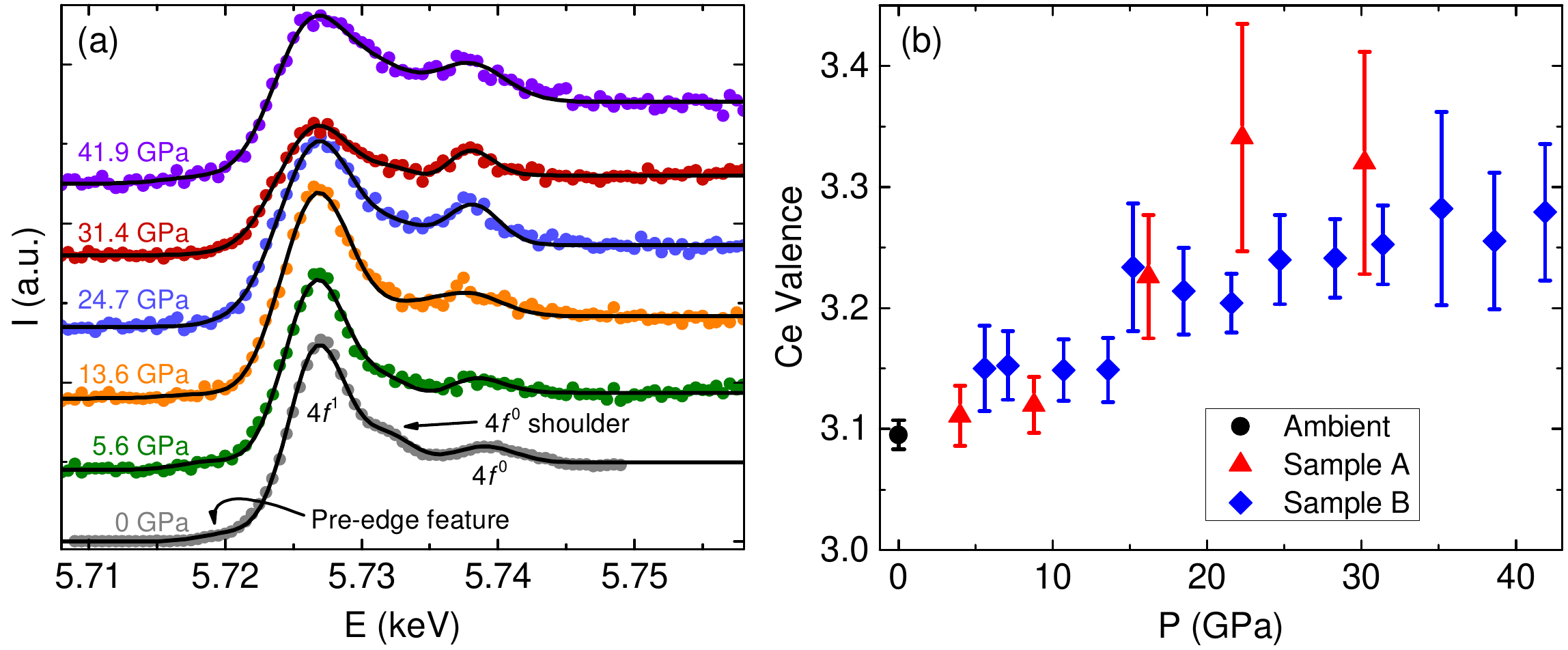}
    \caption{(a) Emitted intensity (offset) vs. incoming photon energy in the region of the Ce L$_3$ edge from XANES measurements for an ambient pressure sample and high pressure Sample B, with fits shown as black lines. Data were taken at room temperature on two separate occasions at the APS, hence the difference in scanned energy range for the ambient sample. A constant background has been subtracted and the data sets have been scaled to have the same high energy fluorescent intensity. The pre-edge feature, 4$f^1$ edge,  4$f^0$-related shoulder, and 4$f^0$ edge are labeled. Only the edge heights factor into valence calculation. (b) Calculated Ce valence (with error bars) as a function of pressure for the ambient pressure crystal and the two DAC samples.}
    \label{fig:Figure2}
\end{figure*}

Though the Kondo gap transition occurs at low temperature, XANES was done at room temperature for several reasons. The Ce L$_3$ edge has a low absorption energy of about 5.73~keV, and the intensity of these low energy x-rays is further attenuated by the diamonds and gasket. The signal would be even weaker at lower temperatures with the addition of a cryostat. Although early calculations predicted a 0.08 change in Ce valence when cooled to cryogenic temperatures\cite{Sanchez-CastroTHDependence} and specific heat data showed a temperature-dependent localization of 4$f^1$ Ce moments\cite{FiskKondoInsulators}, previous ambient pressure XANES on Ce$_3$Bi$_4$Pt$_3$ showed, within error, no valence change from 300-10~K.\cite{KweiCeValence} More recent calculations also expect a temperature-independent valence\cite{CaoTheory}. This would be in line with other Ce-based compounds that show little change in valence inside or outside of the Kondo regime\cite{YamaokaCeTIn5Valence,TsvyashchenkoCeRu2Valence}. Room temperature measurements therefore maximize signal-to-noise while likely giving a comparable result to what would be obtained at lower temperatures.

Three different single crystal samples were used for valence measurements: one at ambient pressure, and two at high pressure (from different batches) identified as A and B. Figure~2(a) shows the emitted intensity for the ambient sample and Sample B at five higher pressures. Each absorption edge will show a peak at the corresponding edge energy, and then a fluorescent background at higher energy. After subtraction of a constant background, the fit was made with a combination of Gaussian peaks and error functions centered on the energies corresponding to four identified features. The two most relevant are the 4$f^1$ (3+) and 4$f^0$ (4+) emission energies that occur at about 5.728 and 5.738~keV, respectively\cite{BrubakerCeRhIn5Valence, YamaokaCeTIn5Valence, TakahashiCeXANES}. These values changed slightly with pressure, as has been noted in other experiments\cite{JosephCePXAS}.

Also noticeable is a pre-edge bump at about 5.720~keV. While some have interpreted this as the 4$f^2$ (2+) edge, we do not do so for several reasons. For one, compounds composed exclusively of 4+ Ce have nevertheless shown a peak at a similar energy\cite{TakahashiCeXANES}. Furthemore, the presence of this feature is inconsistent across data sets, being almost absent for Sample B and for Sample A (not shown) becoming more prominent with pressure as overall valence increases, making its identification as an absorption edge dubious. We take the view of other authors who identify this feature as a 2\textit{p}-4\textit{f} transition\cite{BrubakerCeRhIn5Valence}. The fourth feature is the shoulder between the 3+ and 4+ valence peaks at around 5.733~keV. This shoulder is thought to be a byproduct of many body interactions associated with the 4$f^0$ state\cite{TakahashiCeXANES}, or a transition from the 2$p_{3/2}$ state into an oxygen orbital in CeO$_2$\cite{ZhangCeNanoparticleXANES}. It may be the case that the small crystals oxidized slightly during exposure to air in the process of cell loading. Sample A was in atmosphere for about four hours, while sample B was exposed for less than one hour. That being said, for Sample B the 13~GPa data were actually taken after the 22~GPa data in an attempt to get finer spacing in that region. In decreasing pressure the valence decreased, showing that the overall increase is indeed from Ce$_3$Bi$_4$Pt$_3$ and not increasing oxidation with time, which is also unlikely since upon pressure application the sample is no longer in contact with air. The consistent behavior between the two samples reinforces this point. The former shows a more prominent shoulder, as does the ambient sample which was not encapsulated during the roughly two hour long collection time. In any case, the feature is small and previous work has shown that it should not be included in valence calculations\cite{KaindlXANES}. Thus we need only compare the heights of the two edges labeled in Fig.~2 for valence determination.

There has been some uncertainty about the ambient pressure Ce valence in this material. The smaller lattice parameter in comparison to isostructural La$_3$Bi$_4$Pt$_3$ and Pr$_3$Bi$_4$Pt$_3$, where the rare earth ion is trivalent, is evidence for the loss of the electron in the outermost (4$f$) shell and therefore also an elevated valence\cite{FiskHybridizationGap}. Additionally, the previous XANES measurement\cite{KweiCeValence} claimed an ambient temperature and pressure valence of 3.10. However, an indirect determination made by inserting the experimental activation gap into the Anderson impurity model\cite{CooleyCe343Pressure} was about 3.02, rising to only 3.08 by 15~GPa.  A fit to our own data gives a value of 3.09 at 0~GPa, very close to previous absorption results and confirming nonintegral valence even before pressure application. With higher pressure, we see relatively consistent results between the two samples, though Sample A has larger error bars. The valence starts out slightly elevated from ambient pressure, and increases to near 3.3 at 42~GPa, with a possible small jump near 15~GPa. This is around the same pressure where XRD may show signs of disruption in the lattice; that being said, in both inelastic and elastic x-ray measurements the change is less than measurement error. There is some disagreement between the two samples, but that is also within the error bars. Comparing to related materials, bulk Ce valence increases by 0.16 up to 2~GPa,\cite{RueffBulkCePressure} and the change is similar in heavy fermion CeCu$_2$Si$_2$ by 7.8~GPa.\cite{RueffCeCu2Si2Pressure} The overall increase here is higher, but requires substantially more pressure. The movement away from a magnetic 3+ Ce valence state indicates further delocalization of the single 4\textit{f} electron, in line with the increase in interactions expected from the smooth unit cell contraction.

\section{\label{sec:Transport}Electrical Transport}

\begin{figure}
    \centering
    \includegraphics[width=0.43\textwidth]{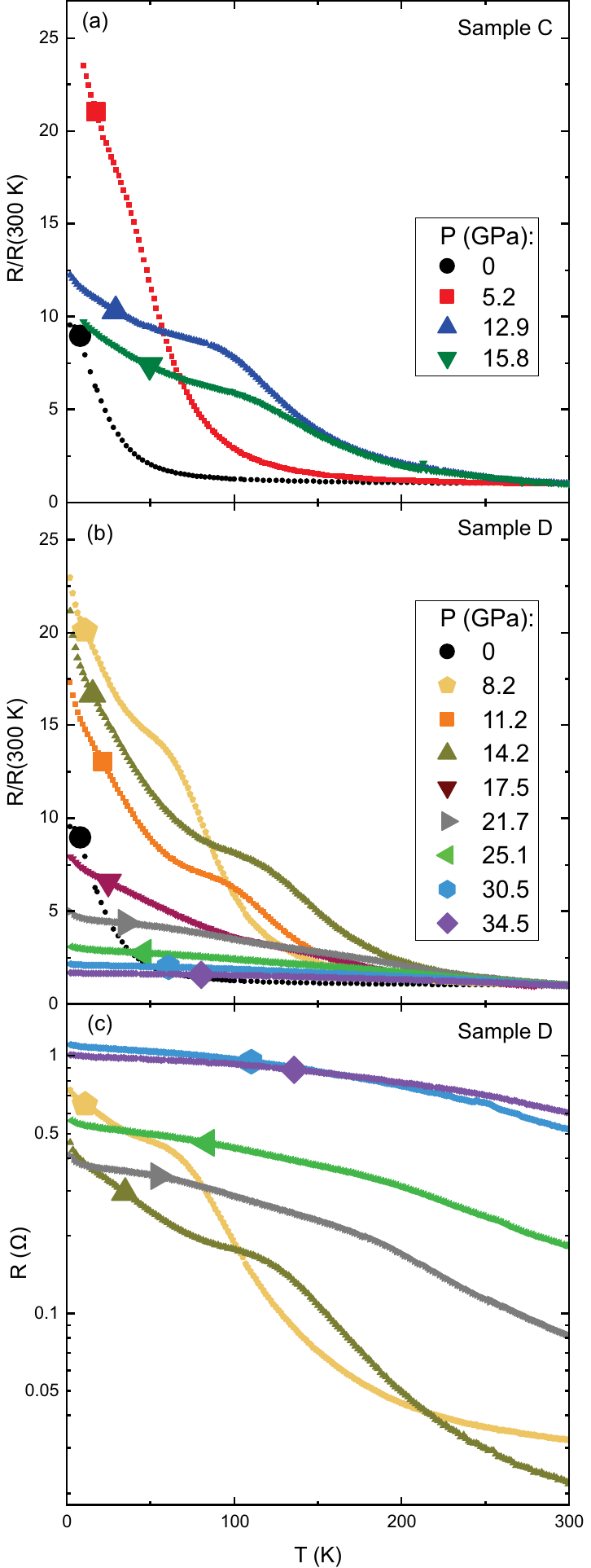}
    \caption{Resistance, scaled to 300~K, as a function of temperature for Ce$_3$Bi$_4$Pt$_3$ Samples (a) C and (b) D at various pressures. The black circular symbols are data from a different ambient pressure sample. (c) Raw resistance values of Sample D data for select pressures, which all had the same wiring configuration. Symbols and colors correspond to the same pressures as in (b).}
    \label{fig:Figure3}
\end{figure}

High pressure resistance measurements were made on two single crystal samples. Crystals grown under the same conditions were also measured at ambient pressure for comparison. Temperature-dependent data are presented in Fig.~3. Where pressures overlap between the two samples, results are very similar and also in line with earlier work up to 14~GPa where hydrostatic pressure was applied with a different method\cite{CooleyCe343Pressure}. There is no indication of superconductivity from Bi flux impurities\cite{LiBiSC}, which affected data in the prior study. The scaled resistivity increase is actually larger for the first few pressure points than at ambient pressure, which was also the case with the previous pressure study\cite{CooleyCe343Pressure}. Nevertheless, the amount by which resistance goes up almost universally gets smaller with pressure for samples in the DAC [Fig.~3(a) and (b)]. For Sample D, R$_{\text{Base}}$/R(300~K) goes from 23 at 8.2~GPa to less than 2 at 34~GPa. The dimensions of Sample D were approximately 75~$\times$~75~$\times$~10~$\upmu$m$^3$, making the resulting estimate for the room temperature resistivity 0.3~m$\Upomega$~cm at 8.2~GPa, comparable to the earlier study in a similar pressure range\cite{CooleyCe343Pressure}. This gives rough confirmation of consistency in sample behavior. That work used an Arrhenius model to estimate the size of the hybridization gap. However, plots of ln(R)~vs.~1/T using our data do not reveal a clear, extended linear region, leading us to conclude that activated behavior is not driving conduction in this compound. The discrepancy with the valence calculated through that gap estimation and our direct measurements are further indication that an activated model does not adequately describe high pressure transport.

The lone distinct feature in R(T) is a hump, indicated in Fig.~3(a) for the 8.2~GPa curve, that increases in temperature with pressure and was noted in earlier high pressure work. This feature does not appear at ambient pressure, where there is a much more divergent and clearly insulating temperature dependence to resistance. However, it may be related to the maximum in ambient pressure magnetic susceptibility\cite{HundleyCe343}. We identify T$_{\text{Hump}}$ by the maximum in $\lvert{}\frac{dR}{dT}\rvert{}$, and plot it alongside the scaled resistance increase in Fig.~4(a).  Up to 20~GPa, it increases slightly sublinearly from 50~K to 200~K. It becomes less prominent with pressure, especially above 20~GPa, as the resistance becomes more and more temperature independent. It is observed up to 212~K at 30.5~GPa but by 34.5~GPa is either too subtle or has moved above 300~K. The width of this hump, quantified by the temperature difference between the local minimum and maximum in the derivative, is consistently 40-50~K.

\begin{figure}
    \centering
    \includegraphics[width=.47\textwidth]{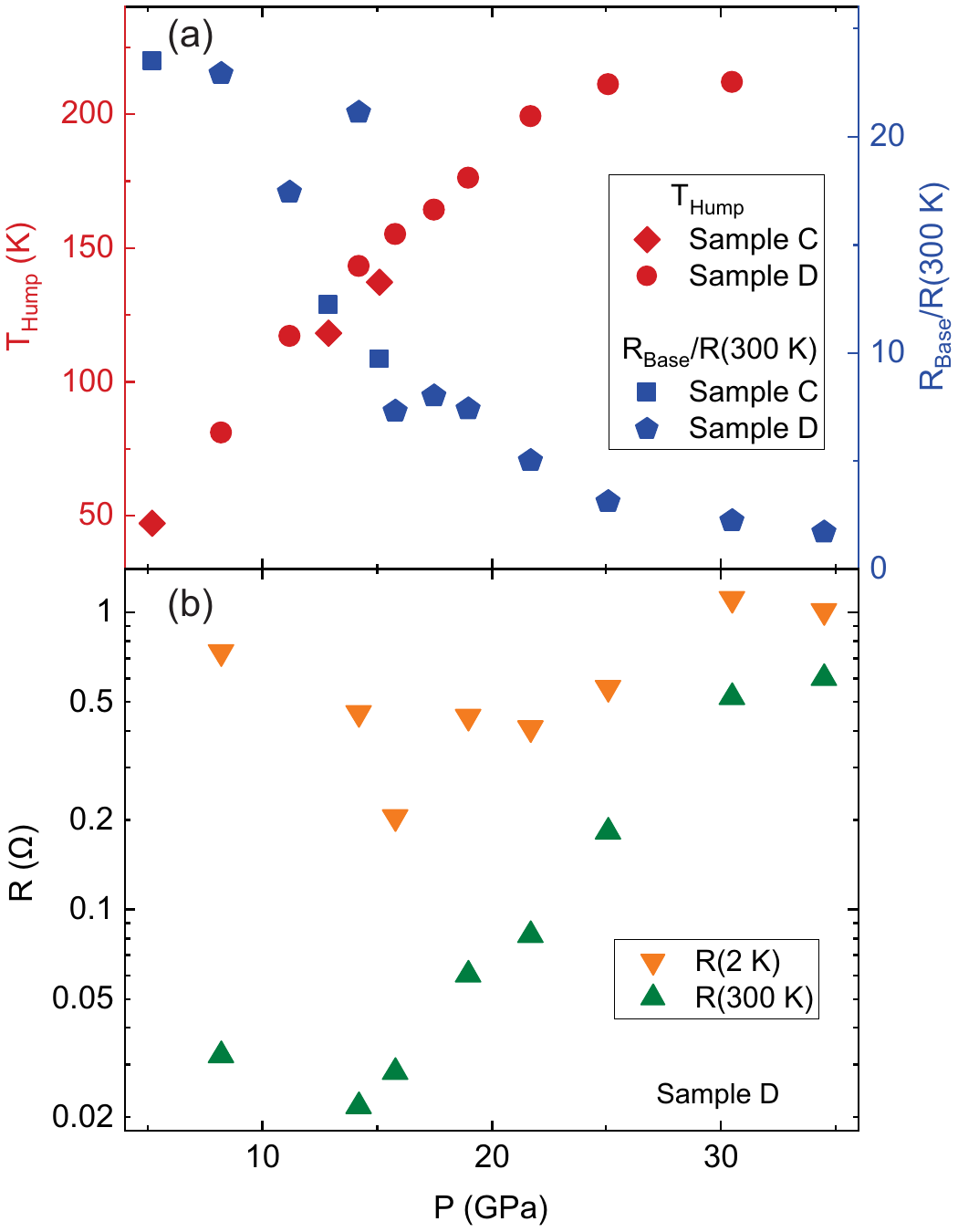}
    \caption{(a) T$_{\text{Hump}}$ (red) and the scaled resistance increase (blue) as a function of pressure for both DAC samples. Base temperature was 2~K for all measurements except Sample C at 5.2~GPa, where it was 10~K. (b) Semilog plot of the resistance at 2~K and 300~K (downward- and upward-facing triangles, respectively) for Sample D, with all measurements done in the same wiring configuration.}
    \label{fig:Figure4}
\end{figure}

A similar feature has been seen in other KIs and metallic Ce-based compounds. Replacing Pt with Pd has led to a diminished resistance increase and hump in Ce$_3$Bi$_4$(Pt$_{1-x}$Pd$_x$)$_3$.\cite{DzsaberCe3PtPd4Bi3} In FeSi, a hump has been linked to the Kondo temperature T$_{\text{K}}$.\cite{SunFeSi} CeRhSb shows a maximum in the magnetic contribution to the resistivity that is attributed to Kondo behavior\cite{UwatokoCeRhSb} and which moves to higher temperatures up to 4.5~GPa. Resistivity maxima in metallic CeCu$_6$\cite{ThompsonCeCu6Pressure}, CePt$_2$Si$_2$\cite{AyacheCePt2Si2Pressure}, Ce$_x$La$_{1-x}$Al$_3$, and Ce$_x$La$_{1-x}$Cu$_2$Si$_2$ all increase in temperature as the unit cell shrinks due to either pressure or increased Ce concentration\cite{BrandtCKS}. In keeping with previous reports, we propose a connection between T$_{\text{Hump}}$ and T$_{\text{K}}$. The movement to higher temperature with pressure represents increasing relevance of hybridization interactions with decreasing unit cell size, as expected and seen in valence data. But while d$\rho$/dT is negative for all temperatures at all pressures, the resistance increase levels off below T$_{\text{Hump}}$, especially at higher pressure. There is also an evident difference in the appearance of the ambient pressure resistance, whose divergence resembles a typical insulator, and all of the DAC measurements. We conclude that instead of reinforcing insulating character, pressure instead leads to metallization and a valence rise, with Kondo physics still relevant. This may be connected to the Weyl-Kondo semimetal state that can arise from Pd substitution\cite{DzsaberCe3PtPd4Bi3,LaiWeylKondoSemimetal}.

Figure~5 shows the transverse magnetoresistance (MR) of an ambient pressure sample and sample D at high pressure. The curves have been symmetrized from positive and negative field data to remove any Hall component, which is harder to avoid given the hexagonal lead geometry in the designer DAC. At ambient pressure there is a miniscule, saturating, positive MR at room temperature and a zero-crossing at intermediate field at low temperature, in line with previous data\cite{HundleyMR, BoebingerMRPulsedField} and theoretical predictions for the KI state\cite{Sanchez-CastroNegativeMR}. However, under pressure this trend is reversed, with MR negative at 300~K positive at 5~K. The resistance change with field also becomes smaller with increasing pressure. At intermediate temperatures it can be positive or negative, and generally shows a larger Hall contribution. A large negative MR at low temperatures has been linked to field-induced gap closure\cite{HundleyMR, BoebingerMRPulsedField, Sanchez-CastroNegativeMR}, an assertion backed by heat capacity work pulsed magnetic fields up to 60~T.\cite{JaimePulsedField} The positive MR at low temperature under pressure then gives evidence for a gap closure before field application, resulting in dominance of a positive magnetoresistance contribution like that seen in many metallic systems. The change in high temperature MR further supports the idea of a global change to material properties, even above the temperature of ambient pressure KI behavior.

\section{\label{sec:Discussion}Discussion}

Ce$_3$Bi$_4$Pt$_3$ shows relatively subtle changes under pressure. There are no dramatic transitions in XRD, and while the valence change of 0.2 is larger than has been seen in bulk Ce or CeCu$_2$Si$_2$,\cite{RueffBulkCePressure, RueffCeCu2Si2Pressure} it comes more gradually, over the course of 40~GPa. Although there is never a resistance decrease on cooling that would signify fully metallic behavior, we assert that pressure suppresses the Kondo insulating gap. Even by 5.2~GPa, R(T) is qualitatively distinct from its ambient pressure form, showing a slowed increase below T$_{\text{Hump}}$ rather than divergence, despite the fact that the scaled resistance increase is actually larger. With further pressure increase there is a nearly monotonic decline in R(2~K)/R(300~K). At 34.5~GPa resistance is nearly temperature independent, certainly not insulating. The development of this state is easiest to track looking at the 14.2, 21.7, and 25.1~GPa curves in Fig.~3(b). The 15-25~GPa range also shows the greatest change in T$_{\text{Hump}}$, R$_{\text{Base}}$/R(300~K), and room temperature resistance. The small jump seen at 15~GPa in the valence, while within error, may also indicate that a more abrupt crossover to a more intermediate valent state takes place in this region.

\begin{figure}
    \centering
    \includegraphics[width=0.43\textwidth]{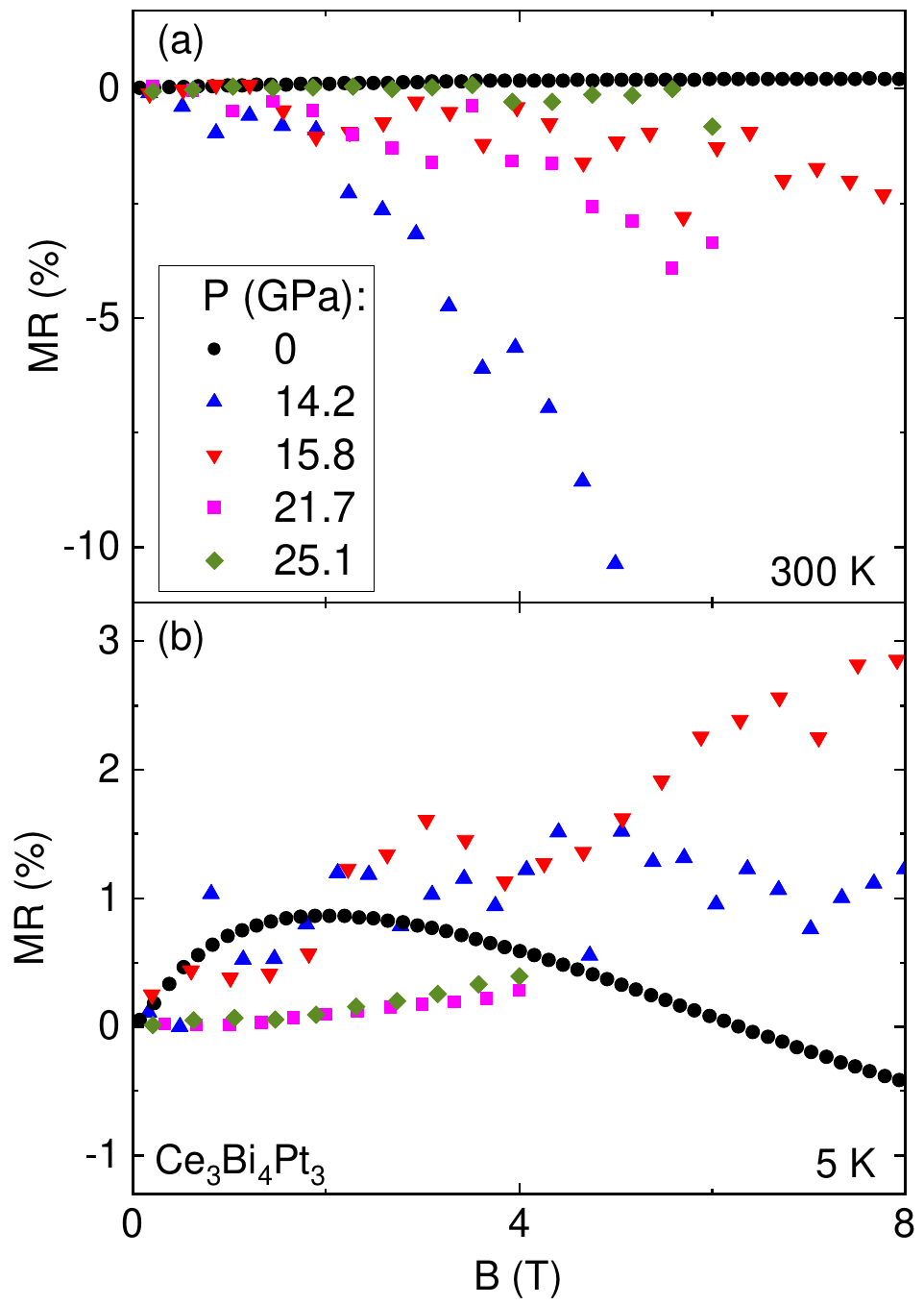}
    \caption{Symmetrized transverse magnetoresistance (as a percentage of zero field resistance) of an ambient pressure crystal and high pressure Sample D at (a) 300~K and (b) 5~K.}
    \label{fig:Figure5}
\end{figure}

It has previously been found that magnetic field\cite{JaimePulsedField,HundleyMR} and La doping\cite{HundleyCe343} close the Ce$_3$Bi$_4$Pt$_3$ Kondo gap. But even at 60~T, the resistance still increases by an order of magnitude from 150~K to 1~K.\cite{BoebingerMRPulsedField}  With 14\% La doping the resistance roughly doubles when cooled to base temperature\cite{HundleyCe343}, a larger increase than we see at the highest achieved pressure. Other Kondo insulators frequently display obvious pressure-induced metallization\cite{BeilleSmB6Gap,UwatokoCeRhSb}, with decreasing resistivities upon cooling. Similarities can be seen to the behavior of Ce$_3$Bi$_4$(Pt$_{1-x}$Pd$_x$)$_3$, which simply shows a flattening of R(T) and decrease in the Kondo gap with isoelectronic substitution, attributed to modification of spin-orbit coupling strength\cite{DzsaberCe3PtPd4Bi3}.

The modest increase in valence over 40~GPa seems divorced from transport behavior, which changes substantially even at the lowest measured pressure. Cooley et al. obtained similar results to ours in resistance measurements up to 14~GPa, but stated that the gap actually increased\cite{CooleyCe343Pressure}. This was based on application of an activated model to the data. However, attempts to use this model with our data did not give satisfactory fits, and we could not find a region where such a model accurately fit the data. The continued flattening out of the resistance at higher pressures than were reached in that study are further evidence for this. Similarly, those authors' attempts to calculate valence from the derived energy gap give substantially lower values than our direct XANES measurements.

The case of Ce$_3$Bi$_4$Pt$_3$ is strong evidence that Kondo insulating behavior can never survive high pressure, even as compression increases hybridization. For the orthorhombic Ce-based KIs like CeNiSn\cite{TakabatakeCeNiSnAnisotropicGap,EchizenCeNiSnPressure}, induced metallic behavior could be ascribed to an anisotropic gap, where the lattice would not shrink isotropically and the Kondo gap would disappear in certain directions. In SmB$_6$\cite{BeilleSmB6Gap,ButchSmB6Valence} and YbB$_{12}$\cite{KayamaYbB12Pressure} the valence change toward a 3+ state is evidence for reduced correlations with pressure. But Ce$_3$Bi$_4$Pt$_3$ is a cubic material that becomes more intermediate valent with pressure, and has a Kondo-related resistance feature that increases in temperature. It would seemingly be a model system for Kondo gap enhancement through compression. Nevertheless, the resistance increase is tempered under pressure, a sign that that there is something lacking in claiming a simple correlation between enhanced hybridization with pressure and a strengthened Kondo gap. Given that the primary difference between a KI and a heavy fermion metal is the position of the Fermi level in the hybridized band structure, shifts of E$_{\text{F}}$ with pressure may also need to be taken into account, especially if the electron number is also changing.

Theoretical work suggested that movement away from integral valence could change Ce$_3$Bi$_4$Pt$_3$ from a weak to strong topological insulator\cite{DzeroTKIs}. The contrast between the sharp increase in room temperature resistance and stability of 2~K resistance [Fig~3(b)] could be interpreted as a residual surface conduction channel taking over at low temperatures as the bulk resistivity increases. Again we make reference to Ce$_3$Bi$_4$(Pt$_{1-x}$Pd$_x$)$_3$, where transition to a ``Kondo semimetal'' may be accompanied by the emergence of topological Weyl points\cite{DzsaberCe3PtPd4Bi3,LaiWeylKondoSemimetal}. Such an assumption could be probed by a nonlocal resistance measurement, like that done to confirm the topological properties of SmB$_6$,\cite{EoCorbino} at high pressure. This is not feasible with diamond anvil cells, but may be possible with other types of high pressure setups.

\section{\label{sec:Conclusion}Conclusion}

We have measured the structure, cerium valence, and resistance of Ce$_3$Bi$_4$Pt$_3$ up to the highest pressures achieved with any Ce-based Kondo insulator. Transport measurements show a crossover from Kondo insulating to an almost flat temperature dependence. The suppression of the resistance increase is even greater than has been achieved with nonmagnetic doping or high magnetic fields, where even though the resistance increased with temperature the behavior was considered ``metallic''\cite{BoebingerMRPulsedField}. The unit cell shrinks monotonically with pressure, leading to increased interactions. This is further evidenced by the increase in Ce valence with pressure as the lone Ce 4\textit{f} electron becomes further delocalized and an increase in the temperature of a resistance hump believed to be related to the Kondo temperature. This material presents a model system with which to observe the connection between uniform compression and increased hybridization interactions. The demonstration of gap closure in a cubic, Ce-based Kondo insulator reinforces the apparently universal antagonism of the Kondo gap and high pressure. Future work could help determine whether the changes we have observed with pressure reflect an enhancement of theorized topological properties.

\section{\label{sec:Acknowledgments}Acknowledgments}

We thank Ryan Stillwell and Sam Weir for assistance with DAC preparation. Portions of this work were supported by LDRD under project 18-SI-001 and under the auspices of the U.S. Department of Energy (DOE) by Lawrence Livermore National Laboratory under contract DE-AC52-07NA27344. Further, this material is based upon work supported by the U.S. DOE, Office of Science, Office of Workforce Development for Teachers and Scientists, Office of Science Graduate Student Research (SCGSR) program. The SCGSR program is administered by the Oak Ridge Institute for Science and Education for the DOE under contract no.~DE‐SC0014664. This work was also supported by Air Force Office of Scientific Research award no.~FA9550-14-1-0332, National Science Foundation Division of Materials Research awards nos.~DMR-1610349 and DMR-1609855, and the Gordon and Betty Moore Foundation's EPiQS Initiative through grant no.~GBMF4419. Portions of this work were performed at HPCAT (Sector~16), Advanced Photon Source, Argonne National Laboratory. HPCAT operations are supported by DOE-NNSA's Office of Experimental Sciences. The Advanced Photon Source is a DOE Office of Science User Facility operated for the DOE Office of Science by Argonne National Laboratory under contract no.~DE-AC02-06CH11357.

\bibliography{Ce343Refs}

\begin{thebibliography}{54}%
\makeatletter
\providecommand \@ifxundefined [1]{%
 \@ifx{#1\undefined}
}%
\providecommand \@ifnum [1]{%
 \ifnum #1\expandafter \@firstoftwo
 \else \expandafter \@secondoftwo
 \fi
}%
\providecommand \@ifx [1]{%
 \ifx #1\expandafter \@firstoftwo
 \else \expandafter \@secondoftwo
 \fi
}%
\providecommand \natexlab [1]{#1}%
\providecommand \enquote  [1]{``#1''}%
\providecommand \bibnamefont  [1]{#1}%
\providecommand \bibfnamefont [1]{#1}%
\providecommand \citenamefont [1]{#1}%
\providecommand \href@noop [0]{\@secondoftwo}%
\providecommand \href [0]{\begingroup \@sanitize@url \@href}%
\providecommand \@href[1]{\@@startlink{#1}\@@href}%
\providecommand \@@href[1]{\endgroup#1\@@endlink}%
\providecommand \@sanitize@url [0]{\catcode `\\12\catcode `\$12\catcode
  `\&12\catcode `\#12\catcode `\^12\catcode `\_12\catcode `\%12\relax}%
\providecommand \@@startlink[1]{}%
\providecommand \@@endlink[0]{}%
\providecommand \url  [0]{\begingroup\@sanitize@url \@url }%
\providecommand \@url [1]{\endgroup\@href {#1}{\urlprefix }}%
\providecommand \urlprefix  [0]{URL }%
\providecommand \Eprint [0]{\href }%
\providecommand \doibase [0]{http://dx.doi.org/}%
\providecommand \selectlanguage [0]{\@gobble}%
\providecommand \bibinfo  [0]{\@secondoftwo}%
\providecommand \bibfield  [0]{\@secondoftwo}%
\providecommand \translation [1]{[#1]}%
\providecommand \BibitemOpen [0]{}%
\providecommand \bibitemStop [0]{}%
\providecommand \bibitemNoStop [0]{.\EOS\space}%
\providecommand \EOS [0]{\spacefactor3000\relax}%
\providecommand \BibitemShut  [1]{\csname bibitem#1\endcsname}%
\let\auto@bib@innerbib\@empty
\bibitem [{\citenamefont {Pietrus}\ \emph {et~al.}(2008)\citenamefont
  {Pietrus}, \citenamefont {v.~L\"ohneysen},\ and\ \citenamefont
  {Schlottmann}}]{PietrusLaDoping}%
  \BibitemOpen
  \bibfield  {author} {\bibinfo {author} {\bibfnamefont {T.}~\bibnamefont
  {Pietrus}}, \bibinfo {author} {\bibfnamefont {H.}~\bibnamefont
  {v.~L\"ohneysen}}, \ and\ \bibinfo {author} {\bibfnamefont {P.}~\bibnamefont
  {Schlottmann}},\ }\href {\doibase 10.1103/PhysRevB.77.115134} {\bibfield
  {journal} {\bibinfo  {journal} {Phys. Rev. B}\ }\textbf {\bibinfo {volume}
  {77}},\ \bibinfo {pages} {115134} (\bibinfo {year} {2008})}\BibitemShut
  {NoStop}%
\bibitem [{\citenamefont {Severing}\ \emph {et~al.}(1991)\citenamefont
  {Severing}, \citenamefont {Thompson}, \citenamefont {Canfield}, \citenamefont
  {Fisk},\ and\ \citenamefont {Riseborough}}]{SeveringInelasticNeutron}%
  \BibitemOpen
  \bibfield  {author} {\bibinfo {author} {\bibfnamefont {A.}~\bibnamefont
  {Severing}}, \bibinfo {author} {\bibfnamefont {J.~D.}\ \bibnamefont
  {Thompson}}, \bibinfo {author} {\bibfnamefont {P.~C.}\ \bibnamefont
  {Canfield}}, \bibinfo {author} {\bibfnamefont {Z.}~\bibnamefont {Fisk}}, \
  and\ \bibinfo {author} {\bibfnamefont {P.}~\bibnamefont {Riseborough}},\
  }\href {\doibase 10.1103/PhysRevB.44.6832} {\bibfield  {journal} {\bibinfo
  {journal} {Phys. Rev. B}\ }\textbf {\bibinfo {volume} {44}},\ \bibinfo
  {pages} {6832} (\bibinfo {year} {1991})}\BibitemShut {NoStop}%
\bibitem [{\citenamefont {Wakeham}\ \emph {et~al.}(2016)\citenamefont
  {Wakeham}, \citenamefont {Rosa}, \citenamefont {Wang}, \citenamefont {Kang},
  \citenamefont {Fisk}, \citenamefont {Ronning},\ and\ \citenamefont
  {Thompson}}]{WakehamTKIs}%
  \BibitemOpen
  \bibfield  {author} {\bibinfo {author} {\bibfnamefont {N.}~\bibnamefont
  {Wakeham}}, \bibinfo {author} {\bibfnamefont {P.~F.~S.}\ \bibnamefont
  {Rosa}}, \bibinfo {author} {\bibfnamefont {Y.~Q.}\ \bibnamefont {Wang}},
  \bibinfo {author} {\bibfnamefont {M.}~\bibnamefont {Kang}}, \bibinfo {author}
  {\bibfnamefont {Z.}~\bibnamefont {Fisk}}, \bibinfo {author} {\bibfnamefont
  {F.}~\bibnamefont {Ronning}}, \ and\ \bibinfo {author} {\bibfnamefont
  {J.~D.}\ \bibnamefont {Thompson}},\ }\href {\doibase
  10.1103/PhysRevB.94.035127} {\bibfield  {journal} {\bibinfo  {journal} {Phys.
  Rev. B}\ }\textbf {\bibinfo {volume} {94}},\ \bibinfo {pages} {035127}
  (\bibinfo {year} {2016})}\BibitemShut {NoStop}%
\bibitem [{\citenamefont {Cao}\ \emph {et~al.}(2019)\citenamefont {Cao},
  \citenamefont {Zhi},\ and\ \citenamefont {Zhu}}]{CaoTheory}%
  \BibitemOpen
  \bibfield  {author} {\bibinfo {author} {\bibfnamefont {C.}~\bibnamefont
  {Cao}}, \bibinfo {author} {\bibfnamefont {G.-X.}\ \bibnamefont {Zhi}}, \ and\
  \bibinfo {author} {\bibfnamefont {J.-X.}\ \bibnamefont {Zhu}},\ }\href
  {https://arxiv.org/pdf/1904.00675.pdf} {\bibfield  {journal} {\bibinfo
  {journal} {arXiv:1904.00675}\ } (\bibinfo {year} {2019})}\BibitemShut
  {NoStop}%
\bibitem [{\citenamefont {Dzero}\ \emph {et~al.}(2010)\citenamefont {Dzero},
  \citenamefont {Sun}, \citenamefont {Galitski},\ and\ \citenamefont
  {Coleman}}]{DzeroTKIs}%
  \BibitemOpen
  \bibfield  {author} {\bibinfo {author} {\bibfnamefont {M.}~\bibnamefont
  {Dzero}}, \bibinfo {author} {\bibfnamefont {K.}~\bibnamefont {Sun}}, \bibinfo
  {author} {\bibfnamefont {V.}~\bibnamefont {Galitski}}, \ and\ \bibinfo
  {author} {\bibfnamefont {P.}~\bibnamefont {Coleman}},\ }\href {\doibase
  10.1103/PhysRevLett.104.106408} {\bibfield  {journal} {\bibinfo  {journal}
  {Phys. Rev. Lett.}\ }\textbf {\bibinfo {volume} {104}},\ \bibinfo {pages}
  {106408} (\bibinfo {year} {2010})}\BibitemShut {NoStop}%
\bibitem [{\citenamefont {Hundley}\ \emph {et~al.}(1993)\citenamefont
  {Hundley}, \citenamefont {Lacerda}, \citenamefont {Canfield}, \citenamefont
  {Thompson},\ and\ \citenamefont {Fisk}}]{HundleyMR}%
  \BibitemOpen
  \bibfield  {author} {\bibinfo {author} {\bibfnamefont {M.}~\bibnamefont
  {Hundley}}, \bibinfo {author} {\bibfnamefont {A.}~\bibnamefont {Lacerda}},
  \bibinfo {author} {\bibfnamefont {P.}~\bibnamefont {Canfield}}, \bibinfo
  {author} {\bibfnamefont {J.}~\bibnamefont {Thompson}}, \ and\ \bibinfo
  {author} {\bibfnamefont {Z.}~\bibnamefont {Fisk}},\ }\href {\doibase
  10.1016/0921-4526(93)90593-U} {\bibfield  {journal} {\bibinfo  {journal}
  {Physica B}\ }\textbf {\bibinfo {volume} {186}},\ \bibinfo {pages} {425}
  (\bibinfo {year} {1993})}\BibitemShut {NoStop}%
\bibitem [{\citenamefont {Jaime}\ \emph {et~al.}(2000)\citenamefont {Jaime},
  \citenamefont {Movshovich}, \citenamefont {Stewart}, \citenamefont
  {Beyermann}, \citenamefont {Gomez~Berisso}, \citenamefont {Hundley},
  \citenamefont {Canfield},\ and\ \citenamefont {Sarrao}}]{JaimePulsedField}%
  \BibitemOpen
  \bibfield  {author} {\bibinfo {author} {\bibfnamefont {M.}~\bibnamefont
  {Jaime}}, \bibinfo {author} {\bibfnamefont {R.}~\bibnamefont {Movshovich}},
  \bibinfo {author} {\bibfnamefont {G.~R.}\ \bibnamefont {Stewart}}, \bibinfo
  {author} {\bibfnamefont {W.~P.}\ \bibnamefont {Beyermann}}, \bibinfo {author}
  {\bibfnamefont {M.}~\bibnamefont {Gomez~Berisso}}, \bibinfo {author}
  {\bibfnamefont {M.~F.}\ \bibnamefont {Hundley}}, \bibinfo {author}
  {\bibfnamefont {P.~C.}\ \bibnamefont {Canfield}}, \ and\ \bibinfo {author}
  {\bibfnamefont {J.~L.}\ \bibnamefont {Sarrao}},\ }\href {\doibase
  10.1038/35012027} {\bibfield  {journal} {\bibinfo  {journal} {Nature}\
  }\textbf {\bibinfo {volume} {405}},\ \bibinfo {pages} {160} (\bibinfo {year}
  {2000})}\BibitemShut {NoStop}%
\bibitem [{\citenamefont {Boebinger}\ \emph {et~al.}(1995)\citenamefont
  {Boebinger}, \citenamefont {Passner}, \citenamefont {Canfield},\ and\
  \citenamefont {Fisk}}]{BoebingerMRPulsedField}%
  \BibitemOpen
  \bibfield  {author} {\bibinfo {author} {\bibfnamefont {G.}~\bibnamefont
  {Boebinger}}, \bibinfo {author} {\bibfnamefont {A.}~\bibnamefont {Passner}},
  \bibinfo {author} {\bibfnamefont {P.}~\bibnamefont {Canfield}}, \ and\
  \bibinfo {author} {\bibfnamefont {Z.}~\bibnamefont {Fisk}},\ }\href {\doibase
  10.1088/1361-648X/aa9e2b} {\bibfield  {journal} {\bibinfo  {journal} {Physica
  B}\ }\textbf {\bibinfo {volume} {211}},\ \bibinfo {pages} {227} (\bibinfo
  {year} {1995})}\BibitemShut {NoStop}%
\bibitem [{\citenamefont {Beille}\ \emph {et~al.}(1983)\citenamefont {Beille},
  \citenamefont {Maple}, \citenamefont {Wittig}, \citenamefont {Fisk},\ and\
  \citenamefont {DeLong}}]{BeilleSmB6Gap}%
  \BibitemOpen
  \bibfield  {author} {\bibinfo {author} {\bibfnamefont {J.}~\bibnamefont
  {Beille}}, \bibinfo {author} {\bibfnamefont {M.~B.}\ \bibnamefont {Maple}},
  \bibinfo {author} {\bibfnamefont {J.}~\bibnamefont {Wittig}}, \bibinfo
  {author} {\bibfnamefont {Z.}~\bibnamefont {Fisk}}, \ and\ \bibinfo {author}
  {\bibfnamefont {L.~E.}\ \bibnamefont {DeLong}},\ }\href {\doibase
  10.1103/PhysRevB.28.7397} {\bibfield  {journal} {\bibinfo  {journal} {Phys.
  Rev. B}\ }\textbf {\bibinfo {volume} {28}},\ \bibinfo {pages} {7397}
  (\bibinfo {year} {1983})}\BibitemShut {NoStop}%
\bibitem [{\citenamefont {Kurisu}\ \emph {et~al.}(1988)\citenamefont {Kurisu},
  \citenamefont {Takabatake},\ and\ \citenamefont {Fujiwara}}]{KurisuCeNiSn}%
  \BibitemOpen
  \bibfield  {author} {\bibinfo {author} {\bibfnamefont {M.}~\bibnamefont
  {Kurisu}}, \bibinfo {author} {\bibfnamefont {T.}~\bibnamefont {Takabatake}},
  \ and\ \bibinfo {author} {\bibfnamefont {H.}~\bibnamefont {Fujiwara}},\
  }\href {\doibase https://doi.org/10.1016/0038-1098(88)90144-5} {\bibfield
  {journal} {\bibinfo  {journal} {Solid State Commun.}\ }\textbf {\bibinfo
  {volume} {68}},\ \bibinfo {pages} {595 } (\bibinfo {year}
  {1988})}\BibitemShut {NoStop}%
\bibitem [{\citenamefont {Uwatoko}\ \emph {et~al.}(1996)\citenamefont
  {Uwatoko}, \citenamefont {Ishii}, \citenamefont {Oomi}, \citenamefont
  {Takahasi}, \citenamefont {Mori}, \citenamefont {Madru},\ and\ \citenamefont
  {Fisk}}]{UwatokoCeRhSb}%
  \BibitemOpen
  \bibfield  {author} {\bibinfo {author} {\bibfnamefont {Y.}~\bibnamefont
  {Uwatoko}}, \bibinfo {author} {\bibfnamefont {T.}~\bibnamefont {Ishii}},
  \bibinfo {author} {\bibfnamefont {G.}~\bibnamefont {Oomi}}, \bibinfo {author}
  {\bibfnamefont {H.}~\bibnamefont {Takahasi}}, \bibinfo {author}
  {\bibfnamefont {N.}~\bibnamefont {Mori}}, \bibinfo {author} {\bibfnamefont
  {D.}~\bibnamefont {Madru}}, \ and\ \bibinfo {author} {\bibfnamefont
  {Z.}~\bibnamefont {Fisk}},\ }\href {\doibase 10.1143/jpsj.65.27} {\bibfield
  {journal} {\bibinfo  {journal} {J. Phys. Soc. Jpn.}\ }\textbf {\bibinfo
  {volume} {65}},\ \bibinfo {pages} {27} (\bibinfo {year} {1996})}\BibitemShut
  {NoStop}%
\bibitem [{\citenamefont {Cooley}\ \emph {et~al.}(1997)\citenamefont {Cooley},
  \citenamefont {Aronson},\ and\ \citenamefont
  {Canfield}}]{CooleyCe343Pressure}%
  \BibitemOpen
  \bibfield  {author} {\bibinfo {author} {\bibfnamefont {J.~C.}\ \bibnamefont
  {Cooley}}, \bibinfo {author} {\bibfnamefont {M.~C.}\ \bibnamefont {Aronson}},
  \ and\ \bibinfo {author} {\bibfnamefont {P.~C.}\ \bibnamefont {Canfield}},\
  }\href {\doibase 10.1103/PhysRevB.55.7533} {\bibfield  {journal} {\bibinfo
  {journal} {Phys. Rev. B}\ }\textbf {\bibinfo {volume} {55}},\ \bibinfo
  {pages} {7533} (\bibinfo {year} {1997})}\BibitemShut {NoStop}%
\bibitem [{\citenamefont {Hundley}\ \emph {et~al.}(1990)\citenamefont
  {Hundley}, \citenamefont {Canfield}, \citenamefont {Thompson}, \citenamefont
  {Fisk},\ and\ \citenamefont {Lawrence}}]{HundleyCe343}%
  \BibitemOpen
  \bibfield  {author} {\bibinfo {author} {\bibfnamefont {M.~F.}\ \bibnamefont
  {Hundley}}, \bibinfo {author} {\bibfnamefont {P.~C.}\ \bibnamefont
  {Canfield}}, \bibinfo {author} {\bibfnamefont {J.~D.}\ \bibnamefont
  {Thompson}}, \bibinfo {author} {\bibfnamefont {Z.}~\bibnamefont {Fisk}}, \
  and\ \bibinfo {author} {\bibfnamefont {J.~M.}\ \bibnamefont {Lawrence}},\
  }\href {\doibase 10.1103/PhysRevB.42.6842} {\bibfield  {journal} {\bibinfo
  {journal} {Phys. Rev. B}\ }\textbf {\bibinfo {volume} {42}},\ \bibinfo
  {pages} {6842} (\bibinfo {year} {1990})}\BibitemShut {NoStop}%
\bibitem [{\citenamefont {Prescher}\ and\ \citenamefont
  {Prakapenka}(2015)}]{PrescherDioptas}%
  \BibitemOpen
  \bibfield  {author} {\bibinfo {author} {\bibfnamefont {C.}~\bibnamefont
  {Prescher}}\ and\ \bibinfo {author} {\bibfnamefont {V.~B.}\ \bibnamefont
  {Prakapenka}},\ }\href {https://doi.org/10.1080/08957959.2015.1059835}
  {\bibfield  {journal} {\bibinfo  {journal} {High Pressure Res.}\ }\textbf
  {\bibinfo {volume} {35}},\ \bibinfo {pages} {223} (\bibinfo {year}
  {2015})}\BibitemShut {NoStop}%
\bibitem [{\citenamefont {Toby}\ and\ \citenamefont
  {Von~Dreele}(2013)}]{TobyGSAS-II}%
  \BibitemOpen
  \bibfield  {author} {\bibinfo {author} {\bibfnamefont {B.~H.}\ \bibnamefont
  {Toby}}\ and\ \bibinfo {author} {\bibfnamefont {R.~B.}\ \bibnamefont
  {Von~Dreele}},\ }\href {\doibase 10.1107/S0021889813003531} {\bibfield
  {journal} {\bibinfo  {journal} {J. Appl. Crystallogr.}\ }\textbf {\bibinfo
  {volume} {46}},\ \bibinfo {pages} {544} (\bibinfo {year} {2013})}\BibitemShut
  {NoStop}%
\bibitem [{\citenamefont {Klotz}\ \emph {et~al.}(2009)\citenamefont {Klotz},
  \citenamefont {Chervin}, \citenamefont {Munsch},\ and\ \citenamefont
  {Marchand}}]{KlotzPressureMedia}%
  \BibitemOpen
  \bibfield  {author} {\bibinfo {author} {\bibfnamefont {S.}~\bibnamefont
  {Klotz}}, \bibinfo {author} {\bibfnamefont {J.-C.}\ \bibnamefont {Chervin}},
  \bibinfo {author} {\bibfnamefont {P.}~\bibnamefont {Munsch}}, \ and\ \bibinfo
  {author} {\bibfnamefont {G.~L.}\ \bibnamefont {Marchand}},\ }\href {\doibase
  10.1088/0022-3727/42/7/075413} {\bibfield  {journal} {\bibinfo  {journal} {J.
  Phys. D: Appl. Phys.}\ }\textbf {\bibinfo {volume} {42}},\ \bibinfo {pages}
  {075413} (\bibinfo {year} {2009})}\BibitemShut {NoStop}%
\bibitem [{\citenamefont {Patterson}\ \emph {et~al.}(2000)\citenamefont
  {Patterson}, \citenamefont {Catledge}, \citenamefont {Vohra}, \citenamefont
  {Akella},\ and\ \citenamefont {Weir}}]{PattersonDesignerDAC}%
  \BibitemOpen
  \bibfield  {author} {\bibinfo {author} {\bibfnamefont {J.~R.}\ \bibnamefont
  {Patterson}}, \bibinfo {author} {\bibfnamefont {S.~A.}\ \bibnamefont
  {Catledge}}, \bibinfo {author} {\bibfnamefont {Y.~K.}\ \bibnamefont {Vohra}},
  \bibinfo {author} {\bibfnamefont {J.}~\bibnamefont {Akella}}, \ and\ \bibinfo
  {author} {\bibfnamefont {S.~T.}\ \bibnamefont {Weir}},\ }\href {\doibase
  10.1103/PhysRevLett.85.5364} {\bibfield  {journal} {\bibinfo  {journal}
  {Phys. Rev. Lett.}\ }\textbf {\bibinfo {volume} {85}},\ \bibinfo {pages}
  {5364} (\bibinfo {year} {2000})}\BibitemShut {NoStop}%
\bibitem [{\citenamefont {Weir}\ \emph {et~al.}(2000)\citenamefont {Weir},
  \citenamefont {Akella}, \citenamefont {Aracne-Ruddle}, \citenamefont
  {Vohra},\ and\ \citenamefont {Catledge}}]{WeirDesignerDAC}%
  \BibitemOpen
  \bibfield  {author} {\bibinfo {author} {\bibfnamefont {S.~T.}\ \bibnamefont
  {Weir}}, \bibinfo {author} {\bibfnamefont {J.}~\bibnamefont {Akella}},
  \bibinfo {author} {\bibfnamefont {C.}~\bibnamefont {Aracne-Ruddle}}, \bibinfo
  {author} {\bibfnamefont {Y.~K.}\ \bibnamefont {Vohra}}, \ and\ \bibinfo
  {author} {\bibfnamefont {S.~A.}\ \bibnamefont {Catledge}},\ }\href {\doibase
  10.1063/1.1326838} {\bibfield  {journal} {\bibinfo  {journal} {Appl. Phys.
  Lett.}\ }\textbf {\bibinfo {volume} {77}},\ \bibinfo {pages} {3400} (\bibinfo
  {year} {2000})}\BibitemShut {NoStop}%
\bibitem [{\citenamefont {Dewaele}\ \emph {et~al.}(2004)\citenamefont
  {Dewaele}, \citenamefont {Loubeyre},\ and\ \citenamefont
  {Mezouar}}]{DewaeleCuEOS}%
  \BibitemOpen
  \bibfield  {author} {\bibinfo {author} {\bibfnamefont {A.}~\bibnamefont
  {Dewaele}}, \bibinfo {author} {\bibfnamefont {P.}~\bibnamefont {Loubeyre}}, \
  and\ \bibinfo {author} {\bibfnamefont {M.}~\bibnamefont {Mezouar}},\ }\href
  {\doibase 10.1103/PhysRevB.70.094112} {\bibfield  {journal} {\bibinfo
  {journal} {Phys. Rev. B}\ }\textbf {\bibinfo {volume} {70}},\ \bibinfo
  {pages} {094112} (\bibinfo {year} {2004})}\BibitemShut {NoStop}%
\bibitem [{\citenamefont {Piermarini}\ \emph {et~al.}(1975)\citenamefont
  {Piermarini}, \citenamefont {Block}, \citenamefont {Barnett},\ and\
  \citenamefont {Forman}}]{PiermariniRubyCalibration}%
  \BibitemOpen
  \bibfield  {author} {\bibinfo {author} {\bibfnamefont {G.~J.}\ \bibnamefont
  {Piermarini}}, \bibinfo {author} {\bibfnamefont {S.}~\bibnamefont {Block}},
  \bibinfo {author} {\bibfnamefont {J.}~\bibnamefont {Barnett}}, \ and\
  \bibinfo {author} {\bibfnamefont {R.}~\bibnamefont {Forman}},\ }\href
  {\doibase 10.1063/1.321957} {\bibfield  {journal} {\bibinfo  {journal} {J.
  Appl. Phys.}\ }\textbf {\bibinfo {volume} {46}},\ \bibinfo {pages} {2774}
  (\bibinfo {year} {1975})}\BibitemShut {NoStop}%
\bibitem [{\citenamefont {Fisk}\ \emph {et~al.}(1995)\citenamefont {Fisk},
  \citenamefont {Sarrao}, \citenamefont {Thompson}, \citenamefont {Mandrus},
  \citenamefont {Hundley}, \citenamefont {Miglori}, \citenamefont {Bucher},
  \citenamefont {Schlesinger}, \citenamefont {Aeppli}, \citenamefont {Bucher},
  \citenamefont {DiTusa}, \citenamefont {Oglesby}, \citenamefont {Ott},
  \citenamefont {Canfield},\ and\ \citenamefont {Brown}}]{FiskKondoInsulators}%
  \BibitemOpen
  \bibfield  {author} {\bibinfo {author} {\bibfnamefont {Z.}~\bibnamefont
  {Fisk}}, \bibinfo {author} {\bibfnamefont {J.}~\bibnamefont {Sarrao}},
  \bibinfo {author} {\bibfnamefont {J.}~\bibnamefont {Thompson}}, \bibinfo
  {author} {\bibfnamefont {D.}~\bibnamefont {Mandrus}}, \bibinfo {author}
  {\bibfnamefont {M.}~\bibnamefont {Hundley}}, \bibinfo {author} {\bibfnamefont
  {A.}~\bibnamefont {Miglori}}, \bibinfo {author} {\bibfnamefont
  {B.}~\bibnamefont {Bucher}}, \bibinfo {author} {\bibfnamefont
  {Z.}~\bibnamefont {Schlesinger}}, \bibinfo {author} {\bibfnamefont
  {G.}~\bibnamefont {Aeppli}}, \bibinfo {author} {\bibfnamefont
  {E.}~\bibnamefont {Bucher}}, \bibinfo {author} {\bibfnamefont
  {J.}~\bibnamefont {DiTusa}}, \bibinfo {author} {\bibfnamefont
  {C.}~\bibnamefont {Oglesby}}, \bibinfo {author} {\bibfnamefont {H.-R.}\
  \bibnamefont {Ott}}, \bibinfo {author} {\bibfnamefont {P.}~\bibnamefont
  {Canfield}}, \ and\ \bibinfo {author} {\bibfnamefont {S.}~\bibnamefont
  {Brown}},\ }\href
  {https://www.sciencedirect.com/science/article/pii/092145269400588M}
  {\bibfield  {journal} {\bibinfo  {journal} {Physica B}\ }\textbf {\bibinfo
  {volume} {206}},\ \bibinfo {pages} {798} (\bibinfo {year}
  {1995})}\BibitemShut {NoStop}%
\bibitem [{\citenamefont {Birch}(1947)}]{BirchEOS}%
  \BibitemOpen
  \bibfield  {author} {\bibinfo {author} {\bibfnamefont {F.}~\bibnamefont
  {Birch}},\ }\href {\doibase 10.1103/PhysRev.71.809} {\bibfield  {journal}
  {\bibinfo  {journal} {Phys. Rev.}\ }\textbf {\bibinfo {volume} {71}},\
  \bibinfo {pages} {809} (\bibinfo {year} {1947})}\BibitemShut {NoStop}%
\bibitem [{\citenamefont {Kwei}\ \emph {et~al.}(1992)\citenamefont {Kwei},
  \citenamefont {Lawrence}, \citenamefont {Canfield}, \citenamefont
  {Beyermann}, \citenamefont {Thompson}, \citenamefont {Fisk}, \citenamefont
  {Lawson},\ and\ \citenamefont {Goldstone}}]{KweiThermalExpansion}%
  \BibitemOpen
  \bibfield  {author} {\bibinfo {author} {\bibfnamefont {G.~H.}\ \bibnamefont
  {Kwei}}, \bibinfo {author} {\bibfnamefont {J.~M.}\ \bibnamefont {Lawrence}},
  \bibinfo {author} {\bibfnamefont {P.~C.}\ \bibnamefont {Canfield}}, \bibinfo
  {author} {\bibfnamefont {W.~P.}\ \bibnamefont {Beyermann}}, \bibinfo {author}
  {\bibfnamefont {J.~D.}\ \bibnamefont {Thompson}}, \bibinfo {author}
  {\bibfnamefont {Z.}~\bibnamefont {Fisk}}, \bibinfo {author} {\bibfnamefont
  {A.~C.}\ \bibnamefont {Lawson}}, \ and\ \bibinfo {author} {\bibfnamefont
  {J.~A.}\ \bibnamefont {Goldstone}},\ }\href {\doibase
  10.1103/PhysRevB.46.8067} {\bibfield  {journal} {\bibinfo  {journal} {Phys.
  Rev. B}\ }\textbf {\bibinfo {volume} {46}},\ \bibinfo {pages} {8067}
  (\bibinfo {year} {1992})}\BibitemShut {NoStop}%
\bibitem [{\citenamefont {Gschneidner}(1990)}]{GschneidnerREMetals}%
  \BibitemOpen
  \bibfield  {author} {\bibinfo {author} {\bibfnamefont {K.~A.}\ \bibnamefont
  {Gschneidner}},\ }\href {\doibase 10.1007/BF03029283} {\bibfield  {journal}
  {\bibinfo  {journal} {Bull. Alloy Phase Diag.}\ }\textbf {\bibinfo {volume}
  {11}},\ \bibinfo {pages} {216} (\bibinfo {year} {1990})}\BibitemShut
  {NoStop}%
\bibitem [{\citenamefont {Kurita}\ \emph {et~al.}(2009)\citenamefont {Kurita},
  \citenamefont {Hedo}, \citenamefont {Koeda}, \citenamefont {Kobayashi},
  \citenamefont {Sato}, \citenamefont {Sugawara},\ and\ \citenamefont
  {Uwatoko}}]{KuritaCeRu4Sb12}%
  \BibitemOpen
  \bibfield  {author} {\bibinfo {author} {\bibfnamefont {N.}~\bibnamefont
  {Kurita}}, \bibinfo {author} {\bibfnamefont {M.}~\bibnamefont {Hedo}},
  \bibinfo {author} {\bibfnamefont {M.}~\bibnamefont {Koeda}}, \bibinfo
  {author} {\bibfnamefont {M.}~\bibnamefont {Kobayashi}}, \bibinfo {author}
  {\bibfnamefont {H.}~\bibnamefont {Sato}}, \bibinfo {author} {\bibfnamefont
  {H.}~\bibnamefont {Sugawara}}, \ and\ \bibinfo {author} {\bibfnamefont
  {Y.}~\bibnamefont {Uwatoko}},\ }\href {\doibase 10.1103/PhysRevB.79.014441}
  {\bibfield  {journal} {\bibinfo  {journal} {Phys. Rev. B}\ }\textbf {\bibinfo
  {volume} {79}},\ \bibinfo {pages} {014441} (\bibinfo {year}
  {2009})}\BibitemShut {NoStop}%
\bibitem [{\citenamefont {\ifmmode~\acute{S}\else \'{S}\fi{}lebarski}\ \emph
  {et~al.}(1998{\natexlab{a}})\citenamefont {\ifmmode~\acute{S}\else
  \'{S}\fi{}lebarski}, \citenamefont {Jezierski}, \citenamefont {Zygmunt},
  \citenamefont {M\"ahl},\ and\ \citenamefont {Neumann}}]{SlebarskiCeZrTiNiSn}%
  \BibitemOpen
  \bibfield  {author} {\bibinfo {author} {\bibfnamefont {A.}~\bibnamefont
  {\ifmmode~\acute{S}\else \'{S}\fi{}lebarski}}, \bibinfo {author}
  {\bibfnamefont {A.}~\bibnamefont {Jezierski}}, \bibinfo {author}
  {\bibfnamefont {A.}~\bibnamefont {Zygmunt}}, \bibinfo {author} {\bibfnamefont
  {S.}~\bibnamefont {M\"ahl}}, \ and\ \bibinfo {author} {\bibfnamefont
  {M.}~\bibnamefont {Neumann}},\ }\href {\doibase 10.1103/PhysRevB.57.9544}
  {\bibfield  {journal} {\bibinfo  {journal} {Phys. Rev. B}\ }\textbf {\bibinfo
  {volume} {57}},\ \bibinfo {pages} {9544} (\bibinfo {year}
  {1998}{\natexlab{a}})}\BibitemShut {NoStop}%
\bibitem [{\citenamefont {\ifmmode~\acute{S}\else \'{S}\fi{}lebarski}\ \emph
  {et~al.}(1998{\natexlab{b}})\citenamefont {\ifmmode~\acute{S}\else
  \'{S}\fi{}lebarski}, \citenamefont {Jezierski}, \citenamefont {M\"ahl},
  \citenamefont {Neumann},\ and\ \citenamefont
  {Borstel}}]{SlebarskiCeImpurities}%
  \BibitemOpen
  \bibfield  {author} {\bibinfo {author} {\bibfnamefont {A.}~\bibnamefont
  {\ifmmode~\acute{S}\else \'{S}\fi{}lebarski}}, \bibinfo {author}
  {\bibfnamefont {A.}~\bibnamefont {Jezierski}}, \bibinfo {author}
  {\bibfnamefont {S.}~\bibnamefont {M\"ahl}}, \bibinfo {author} {\bibfnamefont
  {M.}~\bibnamefont {Neumann}}, \ and\ \bibinfo {author} {\bibfnamefont
  {G.}~\bibnamefont {Borstel}},\ }\href {\doibase 10.1103/PhysRevB.58.4367}
  {\bibfield  {journal} {\bibinfo  {journal} {Phys. Rev. B}\ }\textbf {\bibinfo
  {volume} {58}},\ \bibinfo {pages} {4367} (\bibinfo {year}
  {1998}{\natexlab{b}})}\BibitemShut {NoStop}%
\bibitem [{\citenamefont {Yamaoka}\ \emph {et~al.}(2015)\citenamefont
  {Yamaoka}, \citenamefont {Yamamoto}, \citenamefont {Schwier}, \citenamefont
  {Honda}, \citenamefont {Zekko}, \citenamefont {Ohta}, \citenamefont {Lin},
  \citenamefont {Nakatake}, \citenamefont {Iwasawa}, \citenamefont {Arita},
  \citenamefont {Shimada}, \citenamefont {Hiraoka}, \citenamefont {Ishii},
  \citenamefont {Tsuei},\ and\ \citenamefont {Mizuki}}]{YamaokaCeTIn5Valence}%
  \BibitemOpen
  \bibfield  {author} {\bibinfo {author} {\bibfnamefont {H.}~\bibnamefont
  {Yamaoka}}, \bibinfo {author} {\bibfnamefont {Y.}~\bibnamefont {Yamamoto}},
  \bibinfo {author} {\bibfnamefont {E.~F.}\ \bibnamefont {Schwier}}, \bibinfo
  {author} {\bibfnamefont {F.}~\bibnamefont {Honda}}, \bibinfo {author}
  {\bibfnamefont {Y.}~\bibnamefont {Zekko}}, \bibinfo {author} {\bibfnamefont
  {Y.}~\bibnamefont {Ohta}}, \bibinfo {author} {\bibfnamefont {J.-F.}\
  \bibnamefont {Lin}}, \bibinfo {author} {\bibfnamefont {M.}~\bibnamefont
  {Nakatake}}, \bibinfo {author} {\bibfnamefont {H.}~\bibnamefont {Iwasawa}},
  \bibinfo {author} {\bibfnamefont {M.}~\bibnamefont {Arita}}, \bibinfo
  {author} {\bibfnamefont {K.}~\bibnamefont {Shimada}}, \bibinfo {author}
  {\bibfnamefont {N.}~\bibnamefont {Hiraoka}}, \bibinfo {author} {\bibfnamefont
  {H.}~\bibnamefont {Ishii}}, \bibinfo {author} {\bibfnamefont {K.-D.}\
  \bibnamefont {Tsuei}}, \ and\ \bibinfo {author} {\bibfnamefont
  {J.}~\bibnamefont {Mizuki}},\ }\href {\doibase 10.1103/PhysRevB.92.235110}
  {\bibfield  {journal} {\bibinfo  {journal} {Phys. Rev. B}\ }\textbf {\bibinfo
  {volume} {92}},\ \bibinfo {pages} {235110} (\bibinfo {year}
  {2015})}\BibitemShut {NoStop}%
\bibitem [{\citenamefont {Joseph}\ \emph {et~al.}(2017)\citenamefont {Joseph},
  \citenamefont {Torchio}, \citenamefont {Benndorf}, \citenamefont {Irifune},
  \citenamefont {Shinmei}, \citenamefont {P{\"o}ttgen},\ and\ \citenamefont
  {Zerr}}]{JosephCePXAS}%
  \BibitemOpen
  \bibfield  {author} {\bibinfo {author} {\bibfnamefont {B.}~\bibnamefont
  {Joseph}}, \bibinfo {author} {\bibfnamefont {R.}~\bibnamefont {Torchio}},
  \bibinfo {author} {\bibfnamefont {C.}~\bibnamefont {Benndorf}}, \bibinfo
  {author} {\bibfnamefont {T.}~\bibnamefont {Irifune}}, \bibinfo {author}
  {\bibfnamefont {T.}~\bibnamefont {Shinmei}}, \bibinfo {author} {\bibfnamefont
  {R.}~\bibnamefont {P{\"o}ttgen}}, \ and\ \bibinfo {author} {\bibfnamefont
  {A.}~\bibnamefont {Zerr}},\ }\href {\doibase 10.1039/C7CP03022C} {\bibfield
  {journal} {\bibinfo  {journal} {Phys. Chem. Chem. Phys.}\ }\textbf {\bibinfo
  {volume} {19}},\ \bibinfo {pages} {17526} (\bibinfo {year}
  {2017})}\BibitemShut {NoStop}%
\bibitem [{\citenamefont {Rueff}\ \emph {et~al.}(2011)\citenamefont {Rueff},
  \citenamefont {Raymond}, \citenamefont {Taguchi}, \citenamefont {Sikora},
  \citenamefont {Iti\'e}, \citenamefont {Baudelet}, \citenamefont
  {Braithwaite}, \citenamefont {Knebel},\ and\ \citenamefont
  {Jaccard}}]{RueffCeCu2Si2Pressure}%
  \BibitemOpen
  \bibfield  {author} {\bibinfo {author} {\bibfnamefont {J.-P.}\ \bibnamefont
  {Rueff}}, \bibinfo {author} {\bibfnamefont {S.}~\bibnamefont {Raymond}},
  \bibinfo {author} {\bibfnamefont {M.}~\bibnamefont {Taguchi}}, \bibinfo
  {author} {\bibfnamefont {M.}~\bibnamefont {Sikora}}, \bibinfo {author}
  {\bibfnamefont {J.-P.}\ \bibnamefont {Iti\'e}}, \bibinfo {author}
  {\bibfnamefont {F.}~\bibnamefont {Baudelet}}, \bibinfo {author}
  {\bibfnamefont {D.}~\bibnamefont {Braithwaite}}, \bibinfo {author}
  {\bibfnamefont {G.}~\bibnamefont {Knebel}}, \ and\ \bibinfo {author}
  {\bibfnamefont {D.}~\bibnamefont {Jaccard}},\ }\href {\doibase
  10.1103/PhysRevLett.106.186405} {\bibfield  {journal} {\bibinfo  {journal}
  {Phys. Rev. Lett.}\ }\textbf {\bibinfo {volume} {106}},\ \bibinfo {pages}
  {186405} (\bibinfo {year} {2011})}\BibitemShut {NoStop}%
\bibitem [{\citenamefont {Butch}\ \emph {et~al.}(2016)\citenamefont {Butch},
  \citenamefont {Paglione}, \citenamefont {Chow}, \citenamefont {Xiao},
  \citenamefont {Marianetti}, \citenamefont {Booth},\ and\ \citenamefont
  {Jeffries}}]{ButchSmB6Valence}%
  \BibitemOpen
  \bibfield  {author} {\bibinfo {author} {\bibfnamefont {N.~P.}\ \bibnamefont
  {Butch}}, \bibinfo {author} {\bibfnamefont {J.}~\bibnamefont {Paglione}},
  \bibinfo {author} {\bibfnamefont {P.}~\bibnamefont {Chow}}, \bibinfo {author}
  {\bibfnamefont {Y.}~\bibnamefont {Xiao}}, \bibinfo {author} {\bibfnamefont
  {C.~A.}\ \bibnamefont {Marianetti}}, \bibinfo {author} {\bibfnamefont
  {C.~H.}\ \bibnamefont {Booth}}, \ and\ \bibinfo {author} {\bibfnamefont
  {J.~R.}\ \bibnamefont {Jeffries}},\ }\href {\doibase
  10.1103/PhysRevLett.116.156401} {\bibfield  {journal} {\bibinfo  {journal}
  {Phys. Rev. Lett.}\ }\textbf {\bibinfo {volume} {116}},\ \bibinfo {pages}
  {156401} (\bibinfo {year} {2016})}\BibitemShut {NoStop}%
\bibitem [{\citenamefont {Fuse}\ \emph {et~al.}(2004)\citenamefont {Fuse},
  \citenamefont {Nakamoto}, \citenamefont {Kurisu}, \citenamefont {Ishimatsu},\
  and\ \citenamefont {Tanida}}]{FuseYbValence}%
  \BibitemOpen
  \bibfield  {author} {\bibinfo {author} {\bibfnamefont {A.}~\bibnamefont
  {Fuse}}, \bibinfo {author} {\bibfnamefont {G.}~\bibnamefont {Nakamoto}},
  \bibinfo {author} {\bibfnamefont {M.}~\bibnamefont {Kurisu}}, \bibinfo
  {author} {\bibfnamefont {N.}~\bibnamefont {Ishimatsu}}, \ and\ \bibinfo
  {author} {\bibfnamefont {H.}~\bibnamefont {Tanida}},\ }\href {\doibase
  10.1016/j.jallcom.2003.12.035} {\bibfield  {journal} {\bibinfo  {journal} {J.
  Alloy. Cmpd.}\ }\textbf {\bibinfo {volume} {376}},\ \bibinfo {pages} {34}
  (\bibinfo {year} {2004})}\BibitemShut {NoStop}%
\bibitem [{\citenamefont {Brubaker}\ \emph {et~al.}(2018)\citenamefont
  {Brubaker}, \citenamefont {Stillwell}, \citenamefont {Chow}, \citenamefont
  {Xiao}, \citenamefont {Kenney-Benson}, \citenamefont {Ferry}, \citenamefont
  {Popov}, \citenamefont {Donald}, \citenamefont {S\"oderlind}, \citenamefont
  {Campbell}, \citenamefont {Paglione}, \citenamefont {Huang}, \citenamefont
  {Baumbach}, \citenamefont {Zieve},\ and\ \citenamefont
  {Jeffries}}]{BrubakerYb114}%
  \BibitemOpen
  \bibfield  {author} {\bibinfo {author} {\bibfnamefont {Z.~E.}\ \bibnamefont
  {Brubaker}}, \bibinfo {author} {\bibfnamefont {R.~L.}\ \bibnamefont
  {Stillwell}}, \bibinfo {author} {\bibfnamefont {P.}~\bibnamefont {Chow}},
  \bibinfo {author} {\bibfnamefont {Y.}~\bibnamefont {Xiao}}, \bibinfo {author}
  {\bibfnamefont {C.}~\bibnamefont {Kenney-Benson}}, \bibinfo {author}
  {\bibfnamefont {R.}~\bibnamefont {Ferry}}, \bibinfo {author} {\bibfnamefont
  {D.}~\bibnamefont {Popov}}, \bibinfo {author} {\bibfnamefont {S.~B.}\
  \bibnamefont {Donald}}, \bibinfo {author} {\bibfnamefont {P.}~\bibnamefont
  {S\"oderlind}}, \bibinfo {author} {\bibfnamefont {D.~J.}\ \bibnamefont
  {Campbell}}, \bibinfo {author} {\bibfnamefont {J.}~\bibnamefont {Paglione}},
  \bibinfo {author} {\bibfnamefont {K.}~\bibnamefont {Huang}}, \bibinfo
  {author} {\bibfnamefont {R.~E.}\ \bibnamefont {Baumbach}}, \bibinfo {author}
  {\bibfnamefont {R.~J.}\ \bibnamefont {Zieve}}, \ and\ \bibinfo {author}
  {\bibfnamefont {J.~R.}\ \bibnamefont {Jeffries}},\ }\href {\doibase
  10.1103/PhysRevB.98.214115} {\bibfield  {journal} {\bibinfo  {journal} {Phys.
  Rev. B}\ }\textbf {\bibinfo {volume} {98}},\ \bibinfo {pages} {214115}
  (\bibinfo {year} {2018})}\BibitemShut {NoStop}%
\bibitem [{\citenamefont {Brubaker}\ \emph {et~al.}(2017)\citenamefont
  {Brubaker}, \citenamefont {Stillwell}, \citenamefont {Chow}, \citenamefont
  {Xiao}, \citenamefont {Kenney-Benson}, \citenamefont {Ferry}, \citenamefont
  {Jenei}, \citenamefont {Zieve},\ and\ \citenamefont
  {Jeffries}}]{BrubakerCeRhIn5Valence}%
  \BibitemOpen
  \bibfield  {author} {\bibinfo {author} {\bibfnamefont {Z.}~\bibnamefont
  {Brubaker}}, \bibinfo {author} {\bibfnamefont {R.}~\bibnamefont {Stillwell}},
  \bibinfo {author} {\bibfnamefont {P.}~\bibnamefont {Chow}}, \bibinfo {author}
  {\bibfnamefont {Y.}~\bibnamefont {Xiao}}, \bibinfo {author} {\bibfnamefont
  {C.}~\bibnamefont {Kenney-Benson}}, \bibinfo {author} {\bibfnamefont
  {R.}~\bibnamefont {Ferry}}, \bibinfo {author} {\bibfnamefont
  {Z.}~\bibnamefont {Jenei}}, \bibinfo {author} {\bibfnamefont
  {R.}~\bibnamefont {Zieve}}, \ and\ \bibinfo {author} {\bibfnamefont
  {J.}~\bibnamefont {Jeffries}},\ }\href
  {https://doi.org/10.1088/1361-648X/aa9e2b} {\bibfield  {journal} {\bibinfo
  {journal} {J. Phys.: Condens. Mat.}\ }\textbf {\bibinfo {volume} {30}},\
  \bibinfo {pages} {035601} (\bibinfo {year} {2017})}\BibitemShut {NoStop}%
\bibitem [{\citenamefont {Sanchez-Castro}(1996)}]{Sanchez-CastroTHDependence}%
  \BibitemOpen
  \bibfield  {author} {\bibinfo {author} {\bibfnamefont {C.}~\bibnamefont
  {Sanchez-Castro}},\ }\href {\doibase 10.1080/13642819608239133} {\bibfield
  {journal} {\bibinfo  {journal} {Philos. Mag. B}\ }\textbf {\bibinfo {volume}
  {73}},\ \bibinfo {pages} {525} (\bibinfo {year} {1996})}\BibitemShut
  {NoStop}%
\bibitem [{\citenamefont {Kwei}\ \emph {et~al.}(1994)\citenamefont {Kwei},
  \citenamefont {Lawrence},\ and\ \citenamefont {Canfield}}]{KweiCeValence}%
  \BibitemOpen
  \bibfield  {author} {\bibinfo {author} {\bibfnamefont {G.~H.}\ \bibnamefont
  {Kwei}}, \bibinfo {author} {\bibfnamefont {J.~M.}\ \bibnamefont {Lawrence}},
  \ and\ \bibinfo {author} {\bibfnamefont {P.~C.}\ \bibnamefont {Canfield}},\
  }\href {\doibase 10.1103/PhysRevB.49.14708} {\bibfield  {journal} {\bibinfo
  {journal} {Phys. Rev. B}\ }\textbf {\bibinfo {volume} {49}},\ \bibinfo
  {pages} {14708} (\bibinfo {year} {1994})}\BibitemShut {NoStop}%
\bibitem [{\citenamefont {Tsvyashchenko}\ \emph {et~al.}(2002)\citenamefont
  {Tsvyashchenko}, \citenamefont {Fomicheva}, \citenamefont {Sorokin},
  \citenamefont {Ryasny}, \citenamefont {Komissarova}, \citenamefont
  {Shpinkova}, \citenamefont {Klementiev}, \citenamefont {Kuznetsov},
  \citenamefont {Menushenkov}, \citenamefont {Trofimov}, \citenamefont
  {Primenko},\ and\ \citenamefont {Cortes}}]{TsvyashchenkoCeRu2Valence}%
  \BibitemOpen
  \bibfield  {author} {\bibinfo {author} {\bibfnamefont {A.~V.}\ \bibnamefont
  {Tsvyashchenko}}, \bibinfo {author} {\bibfnamefont {L.~N.}\ \bibnamefont
  {Fomicheva}}, \bibinfo {author} {\bibfnamefont {A.~A.}\ \bibnamefont
  {Sorokin}}, \bibinfo {author} {\bibfnamefont {G.~K.}\ \bibnamefont {Ryasny}},
  \bibinfo {author} {\bibfnamefont {B.~A.}\ \bibnamefont {Komissarova}},
  \bibinfo {author} {\bibfnamefont {L.~G.}\ \bibnamefont {Shpinkova}}, \bibinfo
  {author} {\bibfnamefont {K.~V.}\ \bibnamefont {Klementiev}}, \bibinfo
  {author} {\bibfnamefont {A.~V.}\ \bibnamefont {Kuznetsov}}, \bibinfo {author}
  {\bibfnamefont {A.~P.}\ \bibnamefont {Menushenkov}}, \bibinfo {author}
  {\bibfnamefont {V.~N.}\ \bibnamefont {Trofimov}}, \bibinfo {author}
  {\bibfnamefont {A.~E.}\ \bibnamefont {Primenko}}, \ and\ \bibinfo {author}
  {\bibfnamefont {R.}~\bibnamefont {Cortes}},\ }\href {\doibase
  10.1103/PhysRevB.65.174513} {\bibfield  {journal} {\bibinfo  {journal} {Phys.
  Rev. B}\ }\textbf {\bibinfo {volume} {65}},\ \bibinfo {pages} {174513}
  (\bibinfo {year} {2002})}\BibitemShut {NoStop}%
\bibitem [{\citenamefont {Takahashi}\ \emph {et~al.}(2002)\citenamefont
  {Takahashi}, \citenamefont {Sakami},\ and\ \citenamefont
  {Nomura}}]{TakahashiCeXANES}%
  \BibitemOpen
  \bibfield  {author} {\bibinfo {author} {\bibfnamefont {Y.}~\bibnamefont
  {Takahashi}}, \bibinfo {author} {\bibfnamefont {H.}~\bibnamefont {Sakami}}, \
  and\ \bibinfo {author} {\bibfnamefont {M.}~\bibnamefont {Nomura}},\ }\href
  {\doibase 10.1016/S0003-2670(02)00709-2} {\bibfield  {journal} {\bibinfo
  {journal} {Anal. Chim. Acta}\ }\textbf {\bibinfo {volume} {468}},\ \bibinfo
  {pages} {345} (\bibinfo {year} {2002})}\BibitemShut {NoStop}%
\bibitem [{\citenamefont {Zhang}\ \emph {et~al.}(2001)\citenamefont {Zhang},
  \citenamefont {Wu}, \citenamefont {Liu}, \citenamefont {Hu}, \citenamefont
  {Wu},\ and\ \citenamefont {Ju}}]{ZhangCeNanoparticleXANES}%
  \BibitemOpen
  \bibfield  {author} {\bibinfo {author} {\bibfnamefont {J.}~\bibnamefont
  {Zhang}}, \bibinfo {author} {\bibfnamefont {Z.}~\bibnamefont {Wu}}, \bibinfo
  {author} {\bibfnamefont {T.}~\bibnamefont {Liu}}, \bibinfo {author}
  {\bibfnamefont {T.}~\bibnamefont {Hu}}, \bibinfo {author} {\bibfnamefont
  {Z.}~\bibnamefont {Wu}}, \ and\ \bibinfo {author} {\bibfnamefont
  {X.}~\bibnamefont {Ju}},\ }\href {\doibase 10.1107/S0909049500016022}
  {\bibfield  {journal} {\bibinfo  {journal} {J. Synchrotron Radiat.}\ }\textbf
  {\bibinfo {volume} {8}},\ \bibinfo {pages} {531} (\bibinfo {year}
  {2001})}\BibitemShut {NoStop}%
\bibitem [{\citenamefont {Kaindl}\ \emph {et~al.}(1988)\citenamefont {Kaindl},
  \citenamefont {Schmiester}, \citenamefont {Sampathkumaran},\ and\
  \citenamefont {Wachter}}]{KaindlXANES}%
  \BibitemOpen
  \bibfield  {author} {\bibinfo {author} {\bibfnamefont {G.}~\bibnamefont
  {Kaindl}}, \bibinfo {author} {\bibfnamefont {G.}~\bibnamefont {Schmiester}},
  \bibinfo {author} {\bibfnamefont {E.~V.}\ \bibnamefont {Sampathkumaran}}, \
  and\ \bibinfo {author} {\bibfnamefont {P.}~\bibnamefont {Wachter}},\ }\href
  {\doibase 10.1103/PhysRevB.38.10174} {\bibfield  {journal} {\bibinfo
  {journal} {Phys. Rev. B}\ }\textbf {\bibinfo {volume} {38}},\ \bibinfo
  {pages} {10174} (\bibinfo {year} {1988})}\BibitemShut {NoStop}%
\bibitem [{\citenamefont {Fisk}\ \emph {et~al.}(1992)\citenamefont {Fisk},
  \citenamefont {Canfield}, \citenamefont {Thompson},\ and\ \citenamefont
  {Hundley}}]{FiskHybridizationGap}%
  \BibitemOpen
  \bibfield  {author} {\bibinfo {author} {\bibfnamefont {Z.}~\bibnamefont
  {Fisk}}, \bibinfo {author} {\bibfnamefont {P.}~\bibnamefont {Canfield}},
  \bibinfo {author} {\bibfnamefont {J.}~\bibnamefont {Thompson}}, \ and\
  \bibinfo {author} {\bibfnamefont {M.}~\bibnamefont {Hundley}},\ }\href
  {\doibase 10.1016/0925-8388(92)90333-5} {\bibfield  {journal} {\bibinfo
  {journal} {J. Alloy. Compd.}\ }\textbf {\bibinfo {volume} {181}},\ \bibinfo
  {pages} {369} (\bibinfo {year} {1992})}\BibitemShut {NoStop}%
\bibitem [{\citenamefont {Rueff}\ \emph {et~al.}(2006)\citenamefont {Rueff},
  \citenamefont {Iti\'e}, \citenamefont {Taguchi}, \citenamefont {Hague},
  \citenamefont {Mariot}, \citenamefont {Delaunay}, \citenamefont {Kappler},\
  and\ \citenamefont {Jaouen}}]{RueffBulkCePressure}%
  \BibitemOpen
  \bibfield  {author} {\bibinfo {author} {\bibfnamefont {J.-P.}\ \bibnamefont
  {Rueff}}, \bibinfo {author} {\bibfnamefont {J.-P.}\ \bibnamefont {Iti\'e}},
  \bibinfo {author} {\bibfnamefont {M.}~\bibnamefont {Taguchi}}, \bibinfo
  {author} {\bibfnamefont {C.~F.}\ \bibnamefont {Hague}}, \bibinfo {author}
  {\bibfnamefont {J.-M.}\ \bibnamefont {Mariot}}, \bibinfo {author}
  {\bibfnamefont {R.}~\bibnamefont {Delaunay}}, \bibinfo {author}
  {\bibfnamefont {J.-P.}\ \bibnamefont {Kappler}}, \ and\ \bibinfo {author}
  {\bibfnamefont {N.}~\bibnamefont {Jaouen}},\ }\href {\doibase
  10.1103/PhysRevLett.96.237403} {\bibfield  {journal} {\bibinfo  {journal}
  {Phys. Rev. Lett.}\ }\textbf {\bibinfo {volume} {96}},\ \bibinfo {pages}
  {237403} (\bibinfo {year} {2006})}\BibitemShut {NoStop}%
\bibitem [{\citenamefont {Li}\ \emph {et~al.}(2017)\citenamefont {Li},
  \citenamefont {Wang}, \citenamefont {Zhu},\ and\ \citenamefont
  {Wen}}]{LiBiSC}%
  \BibitemOpen
  \bibfield  {author} {\bibinfo {author} {\bibfnamefont {Y.}~\bibnamefont
  {Li}}, \bibinfo {author} {\bibfnamefont {E.}~\bibnamefont {Wang}}, \bibinfo
  {author} {\bibfnamefont {X.}~\bibnamefont {Zhu}}, \ and\ \bibinfo {author}
  {\bibfnamefont {H.-H.}\ \bibnamefont {Wen}},\ }\href {\doibase
  10.1103/PhysRevB.95.024510} {\bibfield  {journal} {\bibinfo  {journal} {Phys.
  Rev. B}\ }\textbf {\bibinfo {volume} {95}},\ \bibinfo {pages} {024510}
  (\bibinfo {year} {2017})}\BibitemShut {NoStop}%
\bibitem [{\citenamefont {Dzsaber}\ \emph {et~al.}(2017)\citenamefont
  {Dzsaber}, \citenamefont {Prochaska}, \citenamefont {Sidorenko},
  \citenamefont {Eguchi}, \citenamefont {Svagera}, \citenamefont {Waas},
  \citenamefont {Prokofiev}, \citenamefont {Si},\ and\ \citenamefont
  {Paschen}}]{DzsaberCe3PtPd4Bi3}%
  \BibitemOpen
  \bibfield  {author} {\bibinfo {author} {\bibfnamefont {S.}~\bibnamefont
  {Dzsaber}}, \bibinfo {author} {\bibfnamefont {L.}~\bibnamefont {Prochaska}},
  \bibinfo {author} {\bibfnamefont {A.}~\bibnamefont {Sidorenko}}, \bibinfo
  {author} {\bibfnamefont {G.}~\bibnamefont {Eguchi}}, \bibinfo {author}
  {\bibfnamefont {R.}~\bibnamefont {Svagera}}, \bibinfo {author} {\bibfnamefont
  {M.}~\bibnamefont {Waas}}, \bibinfo {author} {\bibfnamefont {A.}~\bibnamefont
  {Prokofiev}}, \bibinfo {author} {\bibfnamefont {Q.}~\bibnamefont {Si}}, \
  and\ \bibinfo {author} {\bibfnamefont {S.}~\bibnamefont {Paschen}},\ }\href
  {\doibase 10.1103/PhysRevLett.118.246601} {\bibfield  {journal} {\bibinfo
  {journal} {Phys. Rev. Lett.}\ }\textbf {\bibinfo {volume} {118}},\ \bibinfo
  {pages} {246601} (\bibinfo {year} {2017})}\BibitemShut {NoStop}%
\bibitem [{\citenamefont {Sun}\ \emph {et~al.}(2014)\citenamefont {Sun},
  \citenamefont {Wei}, \citenamefont {Menzel},\ and\ \citenamefont
  {Steglich}}]{SunFeSi}%
  \BibitemOpen
  \bibfield  {author} {\bibinfo {author} {\bibfnamefont {P.}~\bibnamefont
  {Sun}}, \bibinfo {author} {\bibfnamefont {B.}~\bibnamefont {Wei}}, \bibinfo
  {author} {\bibfnamefont {D.}~\bibnamefont {Menzel}}, \ and\ \bibinfo {author}
  {\bibfnamefont {F.}~\bibnamefont {Steglich}},\ }\href {\doibase
  10.1103/PhysRevB.90.245146} {\bibfield  {journal} {\bibinfo  {journal} {Phys.
  Rev. B}\ }\textbf {\bibinfo {volume} {90}},\ \bibinfo {pages} {245146}
  (\bibinfo {year} {2014})}\BibitemShut {NoStop}%
\bibitem [{\citenamefont {Thompson}\ and\ \citenamefont
  {Fisk}(1985)}]{ThompsonCeCu6Pressure}%
  \BibitemOpen
  \bibfield  {author} {\bibinfo {author} {\bibfnamefont {J.~D.}\ \bibnamefont
  {Thompson}}\ and\ \bibinfo {author} {\bibfnamefont {Z.}~\bibnamefont
  {Fisk}},\ }\href {\doibase 10.1103/PhysRevB.31.389} {\bibfield  {journal}
  {\bibinfo  {journal} {Phys. Rev. B}\ }\textbf {\bibinfo {volume} {31}},\
  \bibinfo {pages} {389} (\bibinfo {year} {1985})}\BibitemShut {NoStop}%
\bibitem [{\citenamefont {Ayache}\ \emph {et~al.}(1987)\citenamefont {Ayache},
  \citenamefont {Beille}, \citenamefont {Bonjour}, \citenamefont {Calemczuk},
  \citenamefont {Creuzet}, \citenamefont {Gignoux}, \citenamefont {Najib},
  \citenamefont {Schmitt}, \citenamefont {Voiron},\ and\ \citenamefont
  {Zerguine}}]{AyacheCePt2Si2Pressure}%
  \BibitemOpen
  \bibfield  {author} {\bibinfo {author} {\bibfnamefont {C.}~\bibnamefont
  {Ayache}}, \bibinfo {author} {\bibfnamefont {J.}~\bibnamefont {Beille}},
  \bibinfo {author} {\bibfnamefont {E.}~\bibnamefont {Bonjour}}, \bibinfo
  {author} {\bibfnamefont {R.}~\bibnamefont {Calemczuk}}, \bibinfo {author}
  {\bibfnamefont {G.}~\bibnamefont {Creuzet}}, \bibinfo {author} {\bibfnamefont
  {D.}~\bibnamefont {Gignoux}}, \bibinfo {author} {\bibfnamefont
  {A.}~\bibnamefont {Najib}}, \bibinfo {author} {\bibfnamefont
  {D.}~\bibnamefont {Schmitt}}, \bibinfo {author} {\bibfnamefont
  {J.}~\bibnamefont {Voiron}}, \ and\ \bibinfo {author} {\bibfnamefont
  {M.}~\bibnamefont {Zerguine}},\ }\href {\doibase
  https://doi.org/10.1016/0304-8853(87)90601-9} {\bibfield  {journal} {\bibinfo
   {journal} {J. Magn. Magn. Mater.}\ }\textbf {\bibinfo {volume} {63-64}},\
  \bibinfo {pages} {329 } (\bibinfo {year} {1987})}\BibitemShut {NoStop}%
\bibitem [{\citenamefont {Brandt}\ and\ \citenamefont
  {Moshchalkov}(1984)}]{BrandtCKS}%
  \BibitemOpen
  \bibfield  {author} {\bibinfo {author} {\bibfnamefont {N.}~\bibnamefont
  {Brandt}}\ and\ \bibinfo {author} {\bibfnamefont {V.}~\bibnamefont
  {Moshchalkov}},\ }\href {https://doi.org/10.1080/00018738400101681}
  {\bibfield  {journal} {\bibinfo  {journal} {Adv. in Phys.}\ }\textbf
  {\bibinfo {volume} {33}},\ \bibinfo {pages} {373} (\bibinfo {year}
  {1984})}\BibitemShut {NoStop}%
\bibitem [{\citenamefont {Lai}\ \emph {et~al.}(2018)\citenamefont {Lai},
  \citenamefont {Grefe}, \citenamefont {Paschen},\ and\ \citenamefont
  {Si}}]{LaiWeylKondoSemimetal}%
  \BibitemOpen
  \bibfield  {author} {\bibinfo {author} {\bibfnamefont {H.-H.}\ \bibnamefont
  {Lai}}, \bibinfo {author} {\bibfnamefont {S.~E.}\ \bibnamefont {Grefe}},
  \bibinfo {author} {\bibfnamefont {S.}~\bibnamefont {Paschen}}, \ and\
  \bibinfo {author} {\bibfnamefont {Q.}~\bibnamefont {Si}},\ }\href {\doibase
  10.1073/pnas.1715851115} {\bibfield  {journal} {\bibinfo  {journal} {Proc.
  Natl. Acad. Sci. U.S.A.}\ }\textbf {\bibinfo {volume} {115}},\ \bibinfo
  {pages} {93} (\bibinfo {year} {2018})}\BibitemShut {NoStop}%
\bibitem [{\citenamefont {Sanchez-Castro}(1994)}]{Sanchez-CastroNegativeMR}%
  \BibitemOpen
  \bibfield  {author} {\bibinfo {author} {\bibfnamefont {C.}~\bibnamefont
  {Sanchez-Castro}},\ }\href {\doibase 10.1103/PhysRevB.49.4421} {\bibfield
  {journal} {\bibinfo  {journal} {Phys. Rev. B}\ }\textbf {\bibinfo {volume}
  {49}},\ \bibinfo {pages} {4421} (\bibinfo {year} {1994})}\BibitemShut
  {NoStop}%
\bibitem [{\citenamefont {Takabatake}\ \emph {et~al.}(1990)\citenamefont
  {Takabatake}, \citenamefont {Teshima}, \citenamefont {Fujii}, \citenamefont
  {Nishigori}, \citenamefont {Suzuki}, \citenamefont {Fujita}, \citenamefont
  {Yamaguchi}, \citenamefont {Sakurai},\ and\ \citenamefont
  {Jaccard}}]{TakabatakeCeNiSnAnisotropicGap}%
  \BibitemOpen
  \bibfield  {author} {\bibinfo {author} {\bibfnamefont {T.}~\bibnamefont
  {Takabatake}}, \bibinfo {author} {\bibfnamefont {F.}~\bibnamefont {Teshima}},
  \bibinfo {author} {\bibfnamefont {H.}~\bibnamefont {Fujii}}, \bibinfo
  {author} {\bibfnamefont {S.}~\bibnamefont {Nishigori}}, \bibinfo {author}
  {\bibfnamefont {T.}~\bibnamefont {Suzuki}}, \bibinfo {author} {\bibfnamefont
  {T.}~\bibnamefont {Fujita}}, \bibinfo {author} {\bibfnamefont
  {Y.}~\bibnamefont {Yamaguchi}}, \bibinfo {author} {\bibfnamefont
  {J.}~\bibnamefont {Sakurai}}, \ and\ \bibinfo {author} {\bibfnamefont
  {D.}~\bibnamefont {Jaccard}},\ }\href {\doibase 10.1103/PhysRevB.41.9607}
  {\bibfield  {journal} {\bibinfo  {journal} {Phys. Rev. B}\ }\textbf {\bibinfo
  {volume} {41}},\ \bibinfo {pages} {9607} (\bibinfo {year}
  {1990})}\BibitemShut {NoStop}%
\bibitem [{\citenamefont {Echizen}\ \emph {et~al.}(2002)\citenamefont
  {Echizen}, \citenamefont {Umeo}, \citenamefont {Igaue},\ and\ \citenamefont
  {Takabatake}}]{EchizenCeNiSnPressure}%
  \BibitemOpen
  \bibfield  {author} {\bibinfo {author} {\bibfnamefont {Y.}~\bibnamefont
  {Echizen}}, \bibinfo {author} {\bibfnamefont {K.}~\bibnamefont {Umeo}},
  \bibinfo {author} {\bibfnamefont {T.}~\bibnamefont {Igaue}}, \ and\ \bibinfo
  {author} {\bibfnamefont {T.}~\bibnamefont {Takabatake}},\ }\href {\doibase
  10.1088/0953-8984/14/20/309} {\bibfield  {journal} {\bibinfo  {journal} {J.
  Phys.: Condens. Mat.}\ }\textbf {\bibinfo {volume} {14}},\ \bibinfo {pages}
  {5145} (\bibinfo {year} {2002})}\BibitemShut {NoStop}%
\bibitem [{\citenamefont {Kayama}\ \emph {et~al.}(2014)\citenamefont {Kayama},
  \citenamefont {Tanaka}, \citenamefont {Miyake}, \citenamefont {Kagayama},
  \citenamefont {Shimizu},\ and\ \citenamefont {Iga}}]{KayamaYbB12Pressure}%
  \BibitemOpen
  \bibfield  {author} {\bibinfo {author} {\bibfnamefont {S.}~\bibnamefont
  {Kayama}}, \bibinfo {author} {\bibfnamefont {S.}~\bibnamefont {Tanaka}},
  \bibinfo {author} {\bibfnamefont {A.}~\bibnamefont {Miyake}}, \bibinfo
  {author} {\bibfnamefont {T.}~\bibnamefont {Kagayama}}, \bibinfo {author}
  {\bibfnamefont {K.}~\bibnamefont {Shimizu}}, \ and\ \bibinfo {author}
  {\bibfnamefont {F.}~\bibnamefont {Iga}},\ }\enquote {\bibinfo {title}
  {Pressure {I}nduced {I}nsulator-to-{M}etal {T}ransition at 170 {GP}a of
  {K}ondo {S}emiconductor {Y}b{B}$_{12}$},}\ in\ \href {\doibase
  10.7566/JPSCP.3.012024} {\emph {\bibinfo {booktitle} {Proceedings of the
  International Conference on Strongly Correlated Electron Systems
  (SCES2013)}}}\ (\bibinfo {year} {2014})\ p.\ \bibinfo {pages}
  {012024}\BibitemShut {NoStop}%
\bibitem [{\citenamefont {Eo}\ \emph {et~al.}(2018)\citenamefont {Eo},
  \citenamefont {Sun}, \citenamefont {Kurdak}, \citenamefont {Kim},\ and\
  \citenamefont {Fisk}}]{EoCorbino}%
  \BibitemOpen
  \bibfield  {author} {\bibinfo {author} {\bibfnamefont {Y.~S.}\ \bibnamefont
  {Eo}}, \bibinfo {author} {\bibfnamefont {K.}~\bibnamefont {Sun}}, \bibinfo
  {author} {\bibfnamefont {{\c{C}}.}~\bibnamefont {Kurdak}}, \bibinfo {author}
  {\bibfnamefont {D.-J.}\ \bibnamefont {Kim}}, \ and\ \bibinfo {author}
  {\bibfnamefont {Z.}~\bibnamefont {Fisk}},\ }\href {\doibase
  10.1103/PhysRevApplied.9.044006} {\bibfield  {journal} {\bibinfo  {journal}
  {Phys. Rev. Applied}\ }\textbf {\bibinfo {volume} {9}},\ \bibinfo {pages}
  {044006} (\bibinfo {year} {2018})}\BibitemShut {NoStop}%
\end{thebibliography}%

\end{document}